\documentclass[]{aa}
\usepackage{xcolor}
\usepackage{array}
\newcolumntype{P}[1]{>{\centering\arraybackslash}p{#1}}
\usepackage{subfig} 
\usepackage{graphicx}
\usepackage{txfonts}
\usepackage{natbib}
\usepackage{upgreek}

\def\HI{{\ion{H}{I}}}

\def\HeII{{\ion{He}{II}}}

\def\Hdue{{H$_2$}}

\def\OIII{{[\ion{O}{III}]}}
\def\OII{{[\ion{O}{II}]}}
\def\OI{{[\ion{O}{I}]}}

\def\SIII{{[\ion{S}{III}]}}
\def\SII{{[\ion{S}{II}]}}

\def\NaD{{\ion{Na}{I}~D}}

\def\NII{{[\ion{N}{II}]}}

\def\FeII{{[\ion{Fe}{II}]}}
\def\MgII{{\ion{Mg}{II}}}
\def\CaII{{\ion{Ca}{II}}}

\newcommand{\kms}{$\,$km$\,$s$^{-1}$}
\newcommand{\ergs}{$\,$erg$\,$s$^{-1}$}

\begin{document}

\title{Probing multi-phase outflows and AGN feedback in compact radio galaxies: the case of PKSB~1934-63}
\titlerunning{PKS1934-63}
\author{F. Santoro \inst{1,2,3}\fnmsep\thanks{email: santoro@astro.rug.nl, f.santoro@sheffield.ac.uk},
             M. Rose\inst{3},
             R. Morganti\inst{1,2},
             C. Tadhunter\inst{3}, 
             T. A. Oosterloo \inst{1,2} and 
             J. Holt \inst{3,4}}

\institute{ASTRON, the Netherlands Institute for Radio Astronomy, PO 2, 7990 AA, Dwingeloo, NL.\and Kapteyn Astronomical Institute, University of Groningen, PO 800, 9700 AV, Groningen, NL.\and Department of Physics and Astronomy, University of Sheffield, Sheffield S3 7RH, UK. \and Leiden Observatory, Leiden University, PO Box 9513, 2300 RA Leiden, NL}

\date{Received 17/04/2018; accepted 22/06/2018}
 
\abstract {Young radio AGN are pivotal for our understanding of many of the still-debated aspects of AGN feedback. In this paper we present a study of the interstellar medium (ISM) in the compact, peaked-spectrum radio galaxy PKS~B1934-63 using X-shooter observations. Most of the warm ionized gas resides within a circum-nuclear disk with a radius of about 200~pc that is likely to constitute the gas reservoir from which the central black hole feeds. On the other hand, we find a biconical outflow of warm ionized gas with an estimated radius of $59\pm12$~pc. This matches the radial extent of the radio source and suggests that the outflow is jet driven. Thanks to the superior wavelength coverage of the data, we can estimate the density of the warm ionized gas using the trans-auroral line technique, and we find that the outflowing gas has remarkably high density, up to $\log n_{\rm e}~(\rm{cm^{-3}})\simeq5.5$. The estimated mass outflow rate is low ($\dot{M}=$10$^{-3}$-10$^{-1}~\rm{M_{\odot}~yr^{-1}}$), and the AGN feedback operates at relatively low efficiency ($\dot{E}/L_{bol}\sim 10^{-4}$-$10^{-3}\%$). 
In addition, optical and near-IR line ratios show that the expansion of the radio source drives fast shocks (with velocities $v_{\rm s}\gtrsim500$~\kms) that ionize and accelerate the outflowing gas. 

At odds with the properties of other compact, peaked-spectrum radio sources hosting warm ionized gas outflows, we do not find signs of kinematically disturbed or outflowing gas in phases colder than the warm ionized gas. We argue that this is due to the young age of our source and thus to the recent nature of the AGN-ISM interaction, and suggest that cold gas forms within the outflowing material and the shock-ionized outflowing gas of PKS~B1934-63 did not have enough time to cool down and accumulate in a colder phase. This scenario is also supported by the multi-phase outflows of other compact and young radio sources in the literature.
}
 

\maketitle
%

\section{Introduction}

The interaction between the energy released by the central active nucleus (AGN) and the host galaxy's interstellar medium (ISM) is particularly prominent in compact and young radio galaxies, and one of the main manifestations of this interaction is visible in the jet-driven gas outflows that extend on scales of galaxy bulges \citep[see, e.g.,][]{1990A&A...231..333F,1994ASPC...54..341F,2000AJ....120.2284A,2002AJ....123.2333O,2006MNRAS.370.1633H,2008MNRAS.387..639H,2015A&A...575A..44G,2015A&A...580A..43G}.
In the context of galaxy evolution, the negative feedback effect that such outflows, and thus AGN, have on the host galaxy has a crucial role in explaining, for example, scaling relations between the central black hole (BH) and its host galaxy properties \citep{1998A&A...331L...1S, 1999MNRAS.308L..39F, 2003ApJ...596L..27K, 2004ApJ...600..580G,2005Natur.433..604D} and the quenching of the star formation \citep{2003ApJ...599...38B, 2006MNRAS.370..645B, 2016A&A...588A..78B} in massive early-type galaxies (ETG).

Compact and young radio galaxies are identified by the (small) size of their radio emission, and based on the properties of their radio spectra, are classified as compact steep spectrum (CSS) or as gigahertz peaked sources (GPS) \citep[e.g.,][]{2009AN....330..193G,1999A&A...345..769M,2003PASA...20...19M}. Many compact radio galaxies show clear signs of the interaction between the expanding radio jets and the surrounding dense and multi-phase ISM, which slows down (or even prevents) the jet expansion \citep[see][and reference therein]{2008A&A...487..885O,2015ApJ...809..168C,2015AJ....149...74T}, in line with simulation predictions \citep[][]{1997ApJ...485..112B,2012ApJ...757..136W,2016AN....337..167W}. 

These newly born AGN inflating their radio lobes into the surrounding ISM give us the unique opportunity to study many aspects of so-called AGN feedback. In particular, they can help us to probe the efficiency of the AGN feedback in different gas phases, and even more, investigate the origin of the cold gas that is often observed in this harsh environment \citep[see, e.g.,][]{2012A&A...541L...7D,2014Natur.511..440T,Oosterloo2017}. Currently, the acceleration mechanism of outflows is uncertain, and these sources are ideal for probing the relevance that shocks have in accelerating and ionizing outflowing gas.

Even though their actual impact is still unclear, ionized gas outflows are commonly found in compact, young radio sources, and they show more extreme features than the outflows in extended radio sources \citep{2008MNRAS.387..639H}.
In the case of the warm ionized gas, one of the important parameters that contributes to the uncertainties in the estimate of the AGN feedback efficiency is the gas electron density $n_{\rm e}$ \citep[see][for a discussion]{2016AN....337..159T,Harrison2018}.  
The classical line ratios used as density diagnostic, such as the \SII$\lambdaup$6717/$\lambdaup$6731\AA\ ratio, give a reliable estimate only for low densities (i.e., $10^{2}<n_{\rm e}<10^{3.5}~\rm{cm^{-3}}$) and saturate in the high-density regime \citep[see][]{2006agna.book.....O}. 
These low densities might not reflect the actual gas properties, especially in the case of compact radio galaxies, and might result in incorrect values for the mass outflow rate and the AGN feedback efficiency. \cite{2011MNRAS.410.1527H} and \cite{Rose2017} make use of the technique based on the \SII\ and \OII\ trans-auroral lines and find that the gas electron density can reach values up to $n_{\rm e}=10^{4-5}~\rm{cm^{-3}}$ for outflowing gas. 

The occurrence and effects that shocks have on the ISM is also an important, although poorly quantified, component of AGN-driven outflows. It is known that both fast radio jets/lobes and AGN winds are able to shock and accelerate the ambient ISM along their path \citep[see, e.g.,][]{Couto2013,Couto2017}.
Evidence and/or indications of the presence of shocks have often been reported for compact radio sources, and they are usually connected to the expansion of the radio source within the ambient ISM. 
The main evidence for shocks comes from Hubble Space Telescope high-resolution imaging studies of the warm ionized gas \citep{1997ApJS..110..191D,2000AJ....120.2284A,2007ApJ...661...70B,2008A&A...488L..59L}, which in some cases have been complemented by spectroscopic observations of highly broadened emission lines \citep[][]{2008MNRAS.387..639H}. In other cases, kinematically disturbed cooler ISM phases (neutral and molecular) have been found at the location of radio lobes, clearly indicating shock acceleration \citep[e.g.,][]{2000AJ....119.2085O,2013Sci...341.1082M,2014Natur.511..440T,Oosterloo2017}. Mainly due to limitations in the observations, pure spectroscopic evidence of shock-ionized gas is sparse and much harder to find.

Finally, outflows of atomic (\HI) and molecular (warm \Hdue\ and CO) gas have been observed in compact steep-spectrum radio sources such as IC~5063 \citep{2014Natur.511..440T,2015A&A...580A...1M}, PKS~B1345+12 \citep{2013Sci...341.1082M, 2012A&A...541L...7D} and 3C~305 \citep{2005A&A...439..521M}.
However, the origin of the cold outflowing gas it is still not clear in the context of AGN feedback. 
A scenario that is gaining consensus predicts that molecular gas forms in situ, in particular in the post-shock regions of the outflows, rather then surviving the AGN-ISM interaction and being gradually accelerated by entrainment. This is supported by the molecular gas observations of the compact radio source IC~5063 \cite[see][]{2014Natur.511..440T,2015A&A...580A...1M} and by recent simulations by \cite{2017arXiv170603784R}, showing that cold gas can form in the first few $\rm{10}^{5}$~yr after the start of the AGN-ISM interaction. Compact radio sources, with their young age and multi-phase outflows, are ideal objects on which to test this scenario.

In this paper, we use spectroscopic observations of the compact radio source PKS~B1934-63 to characterize the efficiency of the AGN feedback for the warm ionized gas phase and compare it to other classes of objects. 
We also study the presence or relevance of shocks using line ratio diagnostics, and investigate the multi-phase nature of the outflowing gas using the emission of the warm molecular gas. Finally, we perform a first attempt to test the scenario in which cold gas forms within the post-shock regions of outflows by combining our findings with previously published results for other compact radio sources.

\subsection{PKS~B1934-63}

The source PKS~B1934-63 (z=0.1824) is a powerful radio AGN (P$_{\rm 1.4GHz} = 10^{27.2}~\rm{W~Hz}^{-1}$) classified as a GPS by \cite{1997A&A...321..105D}.
It has often been considered as the archetypal GPS source: it is one of the closest and most powerful compact radio sources and was one of the first GPS to be discovered \citep{1963Natur.199..682B}.
Very long baseline interferometry (VLBI) observations resolved the radio source into two components, likely representing the two radio lobes, separated by 131.7$\pm$0.9~pc \citep[][]{2004AJ....127.1977O}.
The kinematic age of the radio source has been estimated to be 1.6$\times10^3$~yr by monitoring the lobe separation over a timescale of about 32 yr \citep{2004AJ....127.1977O}.

The host galaxy of PKS~B1934-63, identified by \cite{1987MNRAS.225..761F}, is an ETG that is undergoing a merger with a companion galaxy located about 9 kpc away.
Optical and IR images revealed the fainter companion together with clear tidal features \citep[][]{1986ApJ...311..526H,2010MNRAS.407.1739I,2011MNRAS.410.1550R}. The optical polarimetry study of \cite{1994MNRAS.271..807T} found polarized light consistent with scattered AGN light, or with nonthermal emission connected to the radio structure. 
 
More recently, \cite{2008MNRAS.387..639H} and \cite{2016MNRAS.459.4259R} studied the conditions of the warm ionized gas in the host galaxy using slit and integral field spectroscopy, respectively.
Both studies reported a broad blueshifted component in the \OIII$\rm{\lambdaup}$5007\AA\ line profile representing outflowing gas, and hints of high gas densities (measured via the classical \SII$\lambdaup$6717/$\lambdaup$6731\AA\ line ratio).  
Moreover, \cite{2016MNRAS.459.4259R} found that the velocity gradient of the outflowing gas component is aligned with the radio jets, and suggested shocks as the mechanism that ionizes the warm ionized gas.

Here, we present long-slit spectroscopic observations of PKS~B1934-63 obtained with the X-shooter instrument \citep{2011A&A...536A.105V} mounted at the VLT. We take advantage of the large wavelength coverage and of the good velocity resolution of the X-shooter data to study the outflowing gas and probe the warm ionized gas electron density via the trans-auroral line technique (Sec.~\ref{DataAnalysisandResults}). In addition, using the spectro-astrometry technique, we study the spatial extent of the different kinematical components of the ionized gas (Sec.~\ref{The properties of the central BH}). This allows us to obtain a better estimate of the mass outflow rate and of the efficiency of the AGN feedback (Sec.~\ref{The warm ionized gas}). 
We also investigate the ionization state of the warm ionized gas using line ratio diagnostic diagrams to probe for the presence of shocks within the outflowing material (Sec.~\ref{ionization_section}). Finally, we study the kinematics of the warm molecular gas and link it to the kinematics of the warm ionized and atomic gas (Sec.~\ref{The H2 warm molecular and neutral gas}). 

Throughout this paper we assume the following cosmology: H$_{0}$=70 \kms Mpc$^{-1}$, $\Omega_0$ = 0.28, and $\Omega_{\lambdaup}$ = 0.72. At the redshift of PKS~B1934-63, 1~arcsec=3.091~kpc.

\begin{figure*}[]
\centering
\includegraphics[width=\hsize, keepaspectratio]{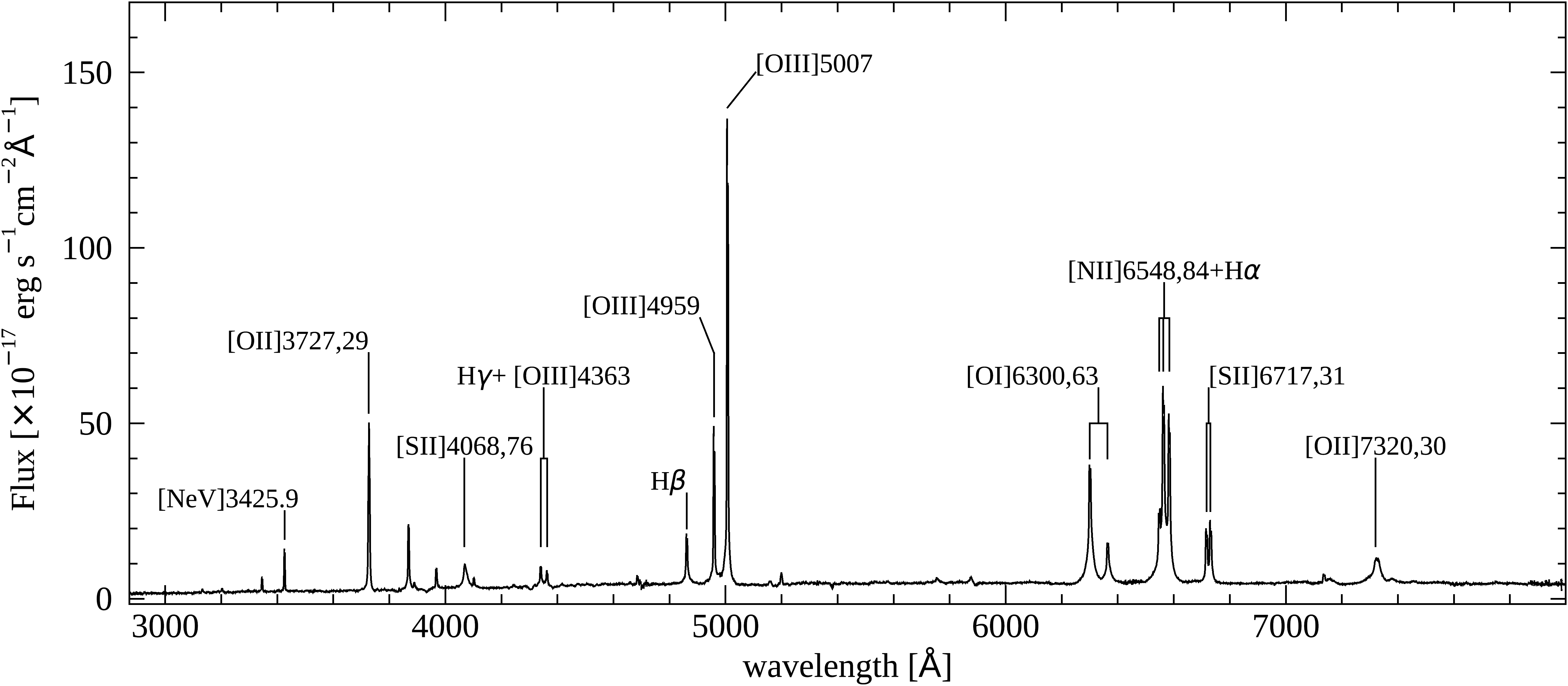}
\caption{UVB+VIS nuclear spectrum of PKS~B1934-63. The main emission lines are indicated. Wavelengths are plotted in \AA,\ and the flux scale is given in units of $10^{17} \rm{erg~s^{-1}cm^{-2}\AA^{-1}}$.} 
\label{panel_spectra}
\end{figure*}

\section{Observations and data reduction}\label{DataReduction}

Observations were carried out with X-shooter at the VLT/UT2 on July 1, 2011, in visitor mode and with a total exposure time of 75 min (i.e., $\rm{10\times450~s}$ for the visual arm (VIS), $\rm{5\times900~s}$ for the ultraviolet-blue arm (UVB), $\rm{15\times300~s}$ for the near-IR arm (NIR)). In order to facilitate sky subtraction, separate exposures were taken with the slit nodded off source. The instrument was used in SLIT mode with 1.6$\times$11 arcsec slit for the UVB arm, 1.5$\times$11 arcsec slit for the VIS arm and 1.5$\times$11 arcsec slit for the NIR arm. The selected slit position angle (PA) was set to be 104 degrees (from north to east), close to the PA of the source's radio axis \citep[i.e., 90 degrees,][]{1989AJ.....98...36T} and including both the PKS~B1934-63 host galaxy and the fainter merging or interacting companion. 

To estimate the seeing, we used three sets of acquisition images taken during the observations, measuring the profiles of seven stars in the images. For each star we extracted a spatial profile, using a mock slit with the same size of the slit used for the actual observations, and we fit it with a Gaussian function. The seeing was then estimated taking the average full width at half-maximum (FWHM) of the fitted 1D profiles. In this way, we obtained a seeing value of 0.97$\pm$0.06 arcsec that takes into account the integration of the seeing profile across the slit in the dispersion direction. The uncertainty in the seeing is the standard error of all the seeing values extracted from the acquisition images.

Standard data reduction was performed using the ESO REFLEX workflow and included bias subtraction, flat fielding, and flux calibration. For each arm, we applied second-order calibrations to the final pipeline products. Residual hot and bad pixels were removed using FIGARO BCCLEAN. Sky subtraction was performed on the slit spectra by extracting an average sky spectrum from the regions of the slit devoid of sources (i.e., the top and bottom part of the slit).
We also performed a telluric absorption-line correction using the integrated spectrum of a standard star observed during the same night.

We derived the average accuracy of the wavelength calibration and the average instrumental width by measuring the line centers and FWHM of sky emission lines, respectively. We found that the wavelength calibration accuracy is 20~\kms, 5~\kms\ , and 3~\kms, while the instrumental width is 90~\kms, 60~\kms\ , and 90~\kms\ for the UVB, VIS, and NIR arms, respectively.  
The relative flux calibration accuracy was estimated to be 10$\%$ taking into account the flux variations of the source calibrated using three different standard stars from the same night. Considering that the radio source is 42.6$\pm$0.3~mas in diameter (i.e., spatially unresolved by the current observations), we used the estimated seeing to set the aperture size and extract the nuclear spectrum of PKS~B1934-63 (see Fig.~\ref{panel_spectra}).

\section{Data analysis and results}\label{DataAnalysisandResults} 

The nuclear spectrum of PKS~B1934-63 (shown in Fig.~\ref{panel_spectra} for the UVB and VIS bands) is extremely rich in emission lines with complex line profiles. The spectrum shows typical features of an high-excitation radio source \citep[HERG,][]{2012MNRAS.421.1569B} and additional strong low-ionization lines such as the \OI, \OII,\ and \SII\ lines. We also detect absorption features for the \MgII$\lambdaup\lambdaup$2796,2804\AA\ in the UVB, and \Hdue\ emission lines in the NIR. 
We find that the strongest emission lines show broad wings and are double peaked. This is discussed in more detail in Sec.~\ref{The emission lines model} together with the line modeling.

\subsection{Redshift and stellar population modeling}\label{z_stellarpop}

Deriving an accurate value for the redshift of the galaxy is essential for the determination of the velocity of any outflowing gas component. The available estimate of the host galaxy systemic velocity is based on bright AGN emission lines \citep{2008MNRAS.387..639H}. These lines are often affected by the complex kinematics of the ionized gas, leading to uncertainties in the derived systemic velocity \citep{2001MNRAS.327..227T, 2009ApJ...698..956C}.

With our current data we were able to estimate the systemic redshift of the galaxy using the \CaII\ K stellar absorption (part of the \CaII$\lambdaup\lambdaup$3933,68\AA\ doublet), which is free from the contamination of gas emission lines.
We fit this line with a Lorentzian function and obtained a redshift $z=0.18240\pm 0.00013$ (the uncertainty on the redshift corresponds to about 40~\kms).

However, the \CaII\ K stellar absorption can potentially include absorption due to the ISM of the host galaxy. To verify that this did not have a significant impact on our redshift estimate, we used the ISM absorption lines of the \MgII$\lambdaup\lambdaup$2795,2802\AA\ doublet.
We found that the width of the \MgII\ absorption lines was significantly lower than that of the \CaII\ K absorption (about seven times lower) and their velocity shift was compatible, within the errors, with the systemic velocity derived from the \CaII\ K absorption (see Appendix~\ref{appendix1} for the details on the fitting procedure).
Therefore, we conclude that the \CaII\ K absorption line profile is mainly related to stellar absorption and that our estimate of the redshift is robust.

We modeled the stellar population in the nuclear spectrum of the host galaxy using STARLIGHT \citep[version 04,][]{2005MNRAS.358..363C} and masking all the emission lines. With the aim of finding a simple model to fit the continuum emission, we used stellar templates with solar metallicity provided by STARLIGHT \citep{2003MNRAS.344.1000B} to model the stellar light. Given that \cite{1994MNRAS.271..807T} detected scattered light from the central AGN, we also introduced a power law ($\rm{F_{\lambdaup}=\lambdaup^{\alpha}}$) in our model to take this into account. 
The best-fit model of the galaxy continuum was chosen based on $\chi^2$ statistics and residual analysis. It includes a 2.5 Gyr old stellar population and a power law with spectral index $\alpha=-0.1$. The redshift derived from the stellar population fit procedure is in line with our redshift estimate using the \CaII\ K absorption line. Our best-fit model is shown in Fig.~\ref{SPfitting} in Appendix~\ref{appendix1} and was subtracted from the nuclear spectrum of PKS~B1934-63 before we modeled the gas emission lines.

\subsection{Kinematic model}\label{The emission lines model}

\begin{figure}[]
\centering
\includegraphics[width=\hsize, keepaspectratio]{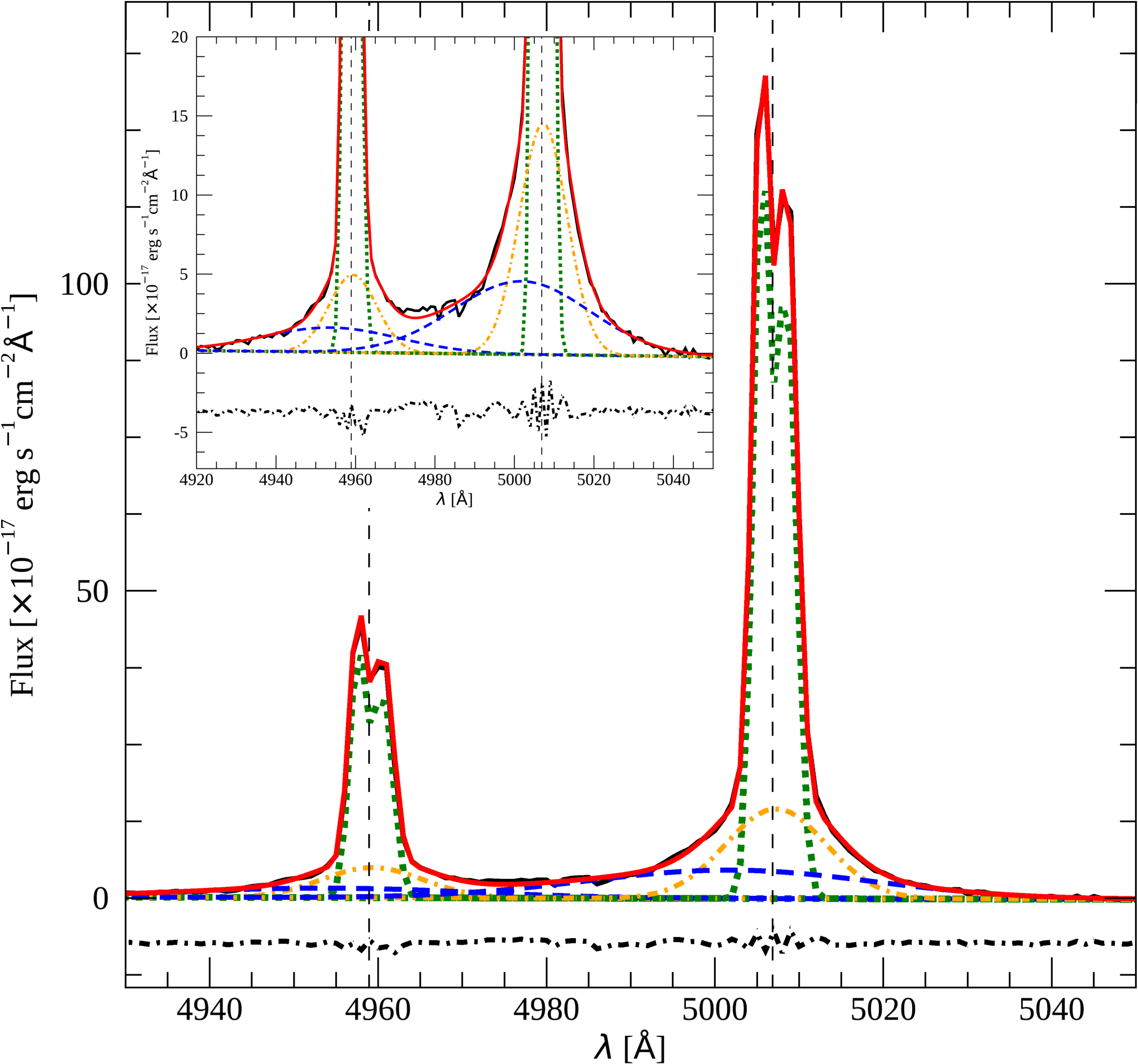}
\caption{  {\rm[O}\,{\rm \scriptsize III}{\rm]}$\lambdaup\lambdaup$4958,5007\AA\ doublet (black solid line) and the {\rm[O}\,{\rm \scriptsize III}{\rm]} model (red solid line). The {\rm[O}\,{\rm \scriptsize III}{\rm]} model includes two narrow components (1N and 2N component, green dotted line), one intermediate component (I component, yellow dot-dashed line), and one very broad component (VB component, blue dashed line). The residuals of the fit are normalized and plotted below the spectrum (black dot-dashed line). The vertical dashed line marks the restframe wavelength of the emission lines. The inset in the top left part of the plot shows a zoom-in of the I and VB components.} 
\label{OIIImodel}
\end{figure}

To obtain a reference model for the forbidden emission lines of the ionized gas, we shifted the spectrum to the galaxy rest-frame and used the \OIII$\lambdaup\lambdaup$4958,5007\AA\ doublet.
Each line of the doublet is double peaked and has broad wings (see Fig.~\ref{OIIImodel}), clearly requiring multiple components to be modeled.
All our fits were performed using Gaussian functions and custom-made IDL routines based on the MPFIT \citep{2009ASPC..411..251M} fitting routine.
For each component of the doublet we forced the width of the Gaussians to be the same. In addition, we fixed their separation (49\AA ) and their relative fluxes (1:3).

The best-fit model of the \OIII$\lambdaup\lambdaup$4958,5007\AA\ doublet (which is called `\OIII\ model') was chosen based on $\chi^2$ statistics and residual minimization. The \OIII\ model, shown in Fig.~\ref{OIIImodel} superposed on the observed lines, includes four components:

\begin{itemize}
\item A narrow redshifted component (1N) with FWHM$_{\rm 1N}$=128$\pm$5~\kms\ and velocity shift  $v_{\rm 1N}$=99$\pm$35~\kms 
\item A narrow blueshifted component (2N) with FWHM$_{\rm 2N}$=104$\pm$4~\kms\ and velocity shift $v_{\rm 2N}$=$-80\pm$35~\kms
\item An intermediate component (I) with FWHM$_{\rm I}$=709$\pm$75~\kms\ and velocity shift $v_{\rm I}$=25$\pm$38~\kms
\item A  very broad blueshifted component (VB) with FWHM$_{\rm VB}$=2035$\pm$207 \kms\ and velocity shift  $v_{\rm VB}$=$-302\pm$112~\kms
\end{itemize}

Velocity shifts were calculated with respect to the systemic velocity of the galaxy (i.e., derived from our redshift estimate), and the FWHM of each component is the intrinsic FWHM, taking into account the instrumental spectral resolution.

\begin{figure}[]
\centering
\includegraphics[width=\hsize, keepaspectratio]{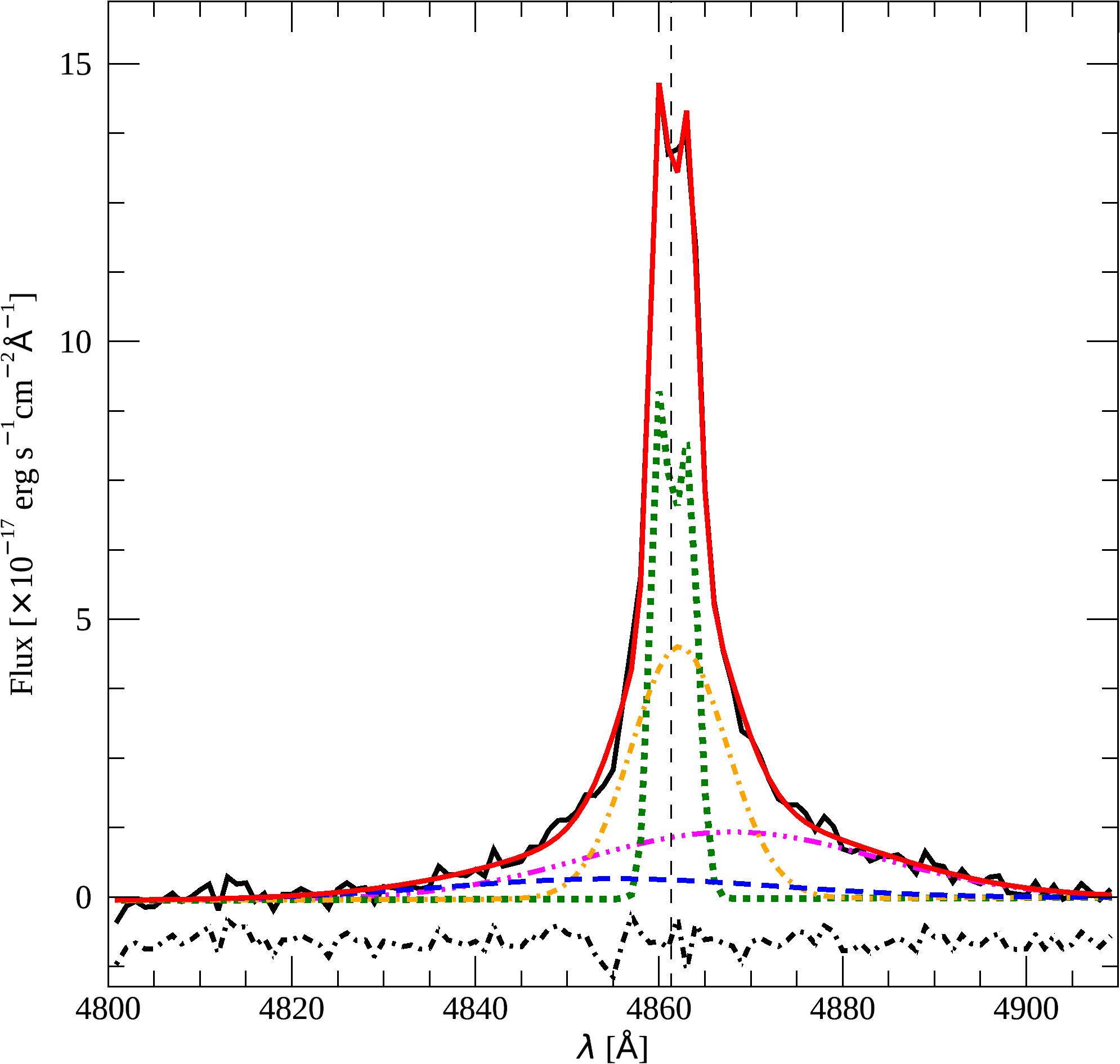}
\caption{H$\beta$ line (black solid line) and the H$\beta$ model (red solid line). The H$\beta$ model includes the four components of the {\rm[O}\,{\rm \scriptsize III}{\rm]} model (see Fig.~\ref{OIIImodel} for a description) and an additional broad redshifted component (magenta triple dot-dashed line). The residuals of the fit are normalized and plotted below the spectrum (black dot-dashed line). The vertical dashed lines marks the restframe wavelength of the emission line.} 
\label{Hbmodel}
\end{figure}

We derived a model for the permitted hydrogen emission lines using the H$\beta$ line. Interestingly, the best-fit model for the H$\beta$ line was consistent with the \OIII\ model, but required an additional redshifted component with $v=398\pm$171~\kms\ and FWHM$=1969\pm$662~\kms. The H$\beta$ model is shown in Fig.~\ref{Hbmodel} superposed on the observed spectrum. 
Nevertheless, this additional component was hard to detect in the other hydrogen emission lines in the nuclear spectrum, mainly due to their weakness (e.g., Pa$\alpha$) or because they blend with other emission lines (e.g., H$\alpha$ and H$\gamma$). 
A possible explanation is that this component is due to light from the central broad line regions of the AGN scattered by an outflowing dusty medium, which can explain the redshift of the component \citep[see, e.g.,][]{diSeregoAlighieri1995,Cimatti1997,Villar2000}. 
  
We found that the \OIII\ model provides a good fit for the \OII$\lambdaup\lambdaup$3726,29\AA, the \OII$\lambdaup\lambdaup$7319,30\AA, the \SII$\lambdaup\lambdaup$4069,76\AA\ + H$\delta$, the \SII$\lambdaup\lambdaup$6717,31\AA, the H$\gamma$ + \OIII$\lambdaup$4363\AA, the \HeII$\lambdaup$4686\AA, the \NII$\lambdaup\lambdaup$6548,84\AA\ + H$\alpha$, the \OI$\lambdaup\lambdaup$6300,63\AA , the \SIII$\lambdaup$9531\AA\ and the Pa$\alpha$ lines. The \OII$\lambdaup\lambdaup$7319,30\AA is a blend of the four {\rm[O}\,{\rm \scriptsize II}{\rm]} lines at 7319, 7320, 7330, and 7331 \AA\ and was fit as a doublet assuming the line centers to be at 7320.1\AA\ and 7330.2\AA\  \citep[based on][]{1979ApJ...229..432S}.
The line fluxes of the four components of each emission line are reported in Table~\ref{Table_linefluxes} in Appendix~\ref{appendix1}.

As has been reported by \cite{2008MNRAS.387..639H} and \cite{2016MNRAS.459.4259R}, the line profile of the \OI$\lambdaup$6300\AA\ line shows a redshifted wing and is different from the rest of the forbidden emission line profiles. Using the bright NIR \SIII$\lambdaup$9531\AA, we found that this is due to contamination by the \SIII$\lambdaup$6312\AA\ line and is not an intrinsic feature of the \OI$\lambdaup$6300\AA\ profile. The \OI$\lambdaup$6300\AA\ line fluxes have been corrected for this when used in line ratio diagnostics (e.g., in Sec.\ref{ionization_section}). In the case of H$\gamma$ + \OIII$\lambdaup$4363\AA,\ we also fit the blend using the H$\beta$ model instead of the \OIII\ model for the H$\gamma$ line. We also note that for \HeII$\lambdaup$4686\AA\ there is no evidence of a very broad component, probably because the line is faint. Additional details on this and on the line fits are reported in Appendix~\ref{appendix1}.  

In the NIR part of the nuclear spectrum of PKS~B1934-63, we detected the \FeII$\rm{\lambdaup}$1.257$\upmu$m and Pa$\beta$ lines, which are used in Sec.~\ref{ionization_section} to study the gas ionization state. The spectrum is noisier and the continuum is not subtracted in the NIR band, which may be the reason that the \OIII\ model did not give a reliable fit for either of these lines. We fit the \FeII$\rm{\lambdaup}$1.257~$\upmu$m line using a single Gaussian function with velocity $v$=$-95\pm$49~\kms\ and FWHM$=547\pm$72~\kms . The Pa$\beta$ line is double peaked, and the best-fitting model included the two narrow components of the \OIII\ model and a third, broader, component centered at $v$=$-233\pm$35~\kms\  with an FWHM$=900\pm$193~\kms .

\subsection{Density diagnostic diagram}\label{DDD_section}

\begin{figure}[t]
\centering
\includegraphics[width=\hsize, keepaspectratio]{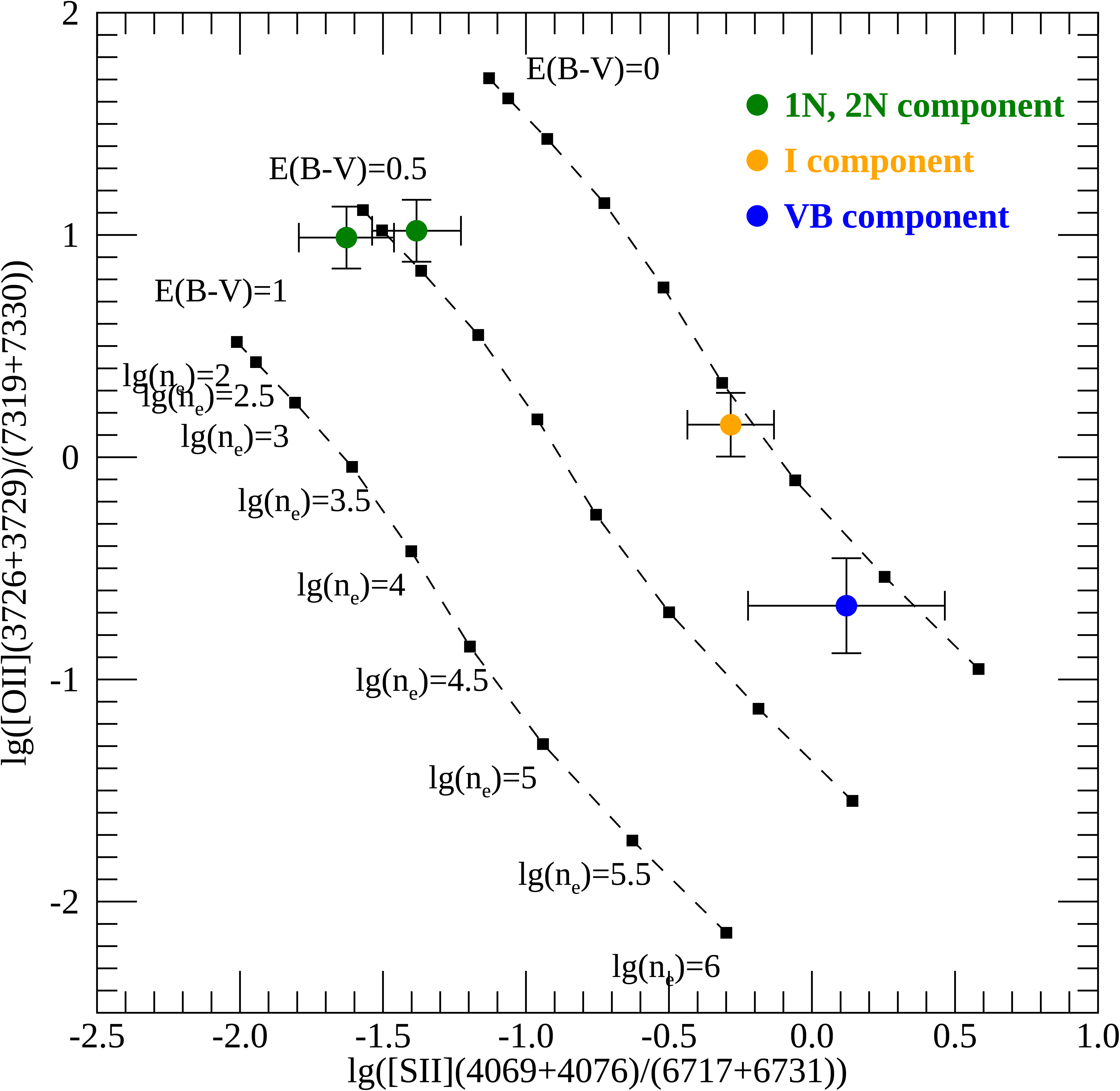}
\caption{ Density diagnostic diagram using the logarithm of the {\rm[O}\,{\rm \scriptsize II}{\rm]} (3727+3729)/(7318+7319+ 7330+7331) and of the {\rm[S}\,{\rm \scriptsize II}{\rm]} (4068+4076)/(6716+6731) line ratios. Each sequence of black squares, joined by the dashed black line, is a sequence of AGN photoionization models with constant power-law index ($\alpha=-$1.5) and ionization parameter ($U$=0.005), created by varying the electron density of the model in the interval $n_{\rm e}=100-10^6~\rm{cm^{-3}}$ (from top left to bottom right) with a step $\Delta$log$_{10}n_{\rm e}=0.5$. The three different sequences in the plot (from top right to bottom left) are associated with $E(B-V)$=0, 0.5 and 1. Green circles represent the narrow components; the intermediate and very broad components are indicated by the golden and blue circle, respectively. Error bars for each point are estimated as described in the text.} 
\label{DDD}
\end{figure}

One of our main goals was to derive a robust estimate of the electron density of the warm ionized gas.  Our observations have a broad wavelength coverage, which enabled us to use the density diagnostic diagram introduced by \cite{2011MNRAS.410.1527H}. This is based on trans-auroral emission lines, and we used it to estimate the electron density of the four different gas components (see Fig.~\ref{DDD}). This diagram uses the \OII\ (3726+3729)/(7319+7330) and  \SII\ (4069+4076)/(6717+6731) line ratios and also provides an estimate of the gas reddening. 
As has been discussed in \cite{2011MNRAS.410.1527H} and \cite{Rose2017}, these diagnostics are sensitive to higher densities than the classical line ratios.

Fig.~\ref{DDD} shows where the four kinematical components of the warm ionized gas are situated in the diagnostic diagram. To derive values for the gas electron density, we have overplotted AGN photoionization models for different reddening factors. The models were produced using the Cloudy \citep[C13.04, ][]{2013RMxAA..49..137F}  photoionization code and the \cite{2000ApJ...533..682C} reddening law. The models shown in Fig.~\ref{DDD} have solar metallicity, a photoionizing continuum with $\alpha=-$1.5 and a ionization parameter $U$=0.005, which reproduces typical conditions of an AGN.

\cite{2011MNRAS.410.1527H} showed that this density diagnostic diagram is not sensitive to the parameters of the AGN photoionization models. This means that the location of the model points in the diagram does not change significantly when the spectral index $\alpha$ of the AGN continuum power law (F$_{\upnu} \propto \upnu^{\alpha}$) and the ionization parameter $U$ are varied.

From the density diagnostic diagram we extracted $\log{n_{e}}(1$N$)~\rm{cm^{-3}}=2.4\pm0.45$, $\log{n_{e}}(2$N$)~\rm{cm^{-3}}=2.7\pm0.45$ for the two narrow components, $\log{n_{e}}($I$)~\rm{cm^{-3}}=4.6\pm0.25$ for the intermediate component, and  $\log{n_{e}}($VB$)~\rm{cm^{-3}}=5.5\pm0.35$ for the very broad component.
The error bars were estimated by summing in quadrature the statistical error from the fitting procedure and the uncertainty in the flux calibration.
It is worth mentioning that the \SII$\lambdaup$6717/$\lambdaup$6731 line ratio, classically used as a density diagnostic, confirmed these results for the narrow components and the intermediate component. The \SII$\lambdaup$6717/$\lambdaup$6731 ratio is 1.10$\pm$0.4 and 1.03$\pm$0.04 for the narrow components and decreases to 0.45$\pm$0.04 for the intermediate component, indicating a density of about 3$\times 10^{2}~$cm$^{-3}$ and higher than $10^4~$cm$^{-3}$ , respectively. The signal-to-noise ratio (S/N) is lower for the very broad component, therefore we could not estimate the density in this way.
The density diagnostic diagram shows that the warm ionized gas spans a significant range of densities from $\sim 3 \times 10^{2.0}~\rm{cm}^{-3}$ up to $10^{5.5}~\rm{cm}^{-3}$, and the higher values are found for the broader components. 

The comparison between the observed points and the sequences of models with different \textit{E(B-V)} values in the density diagnostic diagram allowed us to derive estimates of the reddening of the four kinematical components for the warm ionized gas. We found $E(B-V)_{\rm 1N}=0.52\pm0.12$, $E(B-V)_{\rm 2N}=0.40\pm0.12$ for the narrow components, $E(B-V)_{\rm I}=0.05\pm0.20$ for the intermediate component, and $E(B-V)_{\rm VB}=0.12\pm0.25$ for the very broad component.

We compare these numbers to the reddenings estimated using the classical approach of the hydrogen line ratios (i.e., the Balmer decrement). 
We used the H$\alpha$/H$\beta$ and the Pa$\alpha$/H$\beta$ line ratios and converted them into a color excess \textit{E(B-V)} following the approach of \cite{2013AJ....145...47M} and using the \cite{2000ApJ...533..682C} extinction curve. The errors on the line ratios take into account both the statistical error of the fitting procedure and the uncertainty in the flux calibration.
From the H$\alpha$/H$\beta$ line ratio we obtained $E(B-V)_{\rm 1N}=0.43\pm0.135$, $E(B-V)_{\rm 2N}=0.40\pm0.135$, $E(B-V)_{\rm I}=0.56\pm0.137$ and $E(B-V)_{\rm VB}=0.186\pm0.8$.
From the Pa$\alpha$/H$\beta$ line ratio we obtained $E(B-V)_{\rm 1N}=0.28\pm0.11$, $E(B-V)_{\rm 2N}=0.25\pm0.11$, $E(B-V)_{\rm I}=0.11\pm0.19$ and $E(B-V)_{\rm VB}=0.82\pm0.85$.
Taking into account the uncertainties, these values and the $E(B-V)$ values extracted from trans-auroral lines generally agree well.
The large uncertainties of the classical approach are mainly due to the faintness of the Pa$\alpha$ line (e.g., the difficulty in determining its continuum level) and the complex line blend in which the H$\alpha$ line is included. We thus preferred to adopt the reddening values coming from the density diagnostic diagram, which are based on the strong emission lines.
These values are reported in Table \ref{Table0} and were used in the estimate of the intrinsic \OIII\ and H$\beta$ luminosities of the different gas components.

We found that none of the kinematic components shows high reddening and that the reddening of the intermediate and very broad components is lower than that of the narrow components. This is consistent with the results for some ultraluminous infrared galaxies (ULIRG) in the sample of \cite{Rose2017} but at odds with the results on the compact radio source PKS~B1345+12 obtained by \cite{2011MNRAS.410.1527H}, who found higher reddenings for broader components with the same method. 

\subsection{Radius of the narrow and broad gas components}\label{The radius of the narrow and broad gas components}

To understand whether the warm ionized gas is extended or concentrated in the central regions of the host galaxy, we used the \OIII$\lambdaup$5007\AA\ line and the spatial information contained in the slit spectrum.
We extracted spatial profiles for the warm ionized gas components (i.e., one for the two narrow components and one including the intermediate and the very broad components) and compared them to the seeing of our observations.

To extract these profiles, we collapsed the slit spectrum along the spectral direction over a given velocity range (with respect to the systemic velocity). The selected velocity range for the narrow components was $-226\lesssim v \lesssim$ 279~\kms . For the intermediate and very broad component we took the velocity range $-854\lesssim v \lesssim -348$~\kms . The profile of the galaxy starlight emission was extracted using two windows, one on the red side  ($-2031\lesssim v \lesssim -$3044~\kms ) and one on the blue side ($-5867\lesssim v \lesssim-$4855~\kms ) of the \OIII$\lambdaup\lambdaup$4958,5007\AA\ lines. 

\begin{table*} 
\centering 
\begin{tabular} { p{2.7cm} p{3cm} p{3cm} p{3cm} p{3cm}} 
\hline              
\noalign{\smallskip}
        \textbf{ }                                                                                                       & \textbf{1N Component}                          & \textbf{2N Component}  & \textbf{I Component} & \textbf{VB Component}  \\    
\hline\hline
\noalign{\smallskip}
   $v$~[\kms ] & 99.6$\pm$35.4 & $-80\pm$35.4 & 25$\pm$38.5  & $-302\pm$112 \\
\noalign{\smallskip}
   FWHM~[\kms ] & 128$\pm$5.3 & 104$\pm$4.2 & 709$\pm$75.3 & 2035$\pm$207 \\
\noalign{\smallskip} 
        log~$n_{\rm e} ~\rm{[cm^{-3}]}$         &  $2.4\pm0.45$   & $2.7\pm0.45$ & $4.6\pm0.25$ & $5.5\pm0.35$ \\     
\noalign{\smallskip} 
   $ E(B-V) $ & 0.52$\pm0.125$ &0.4$\pm0.125$ & 0.05$\pm0.2$ & 0.125$\pm0.25$\\
\noalign{\smallskip}
   $L({\rm H\beta})~\rm{[erg~s^{-1}]}$  &  (1.89$\pm$0.20)$\times10^{41}$ &(1.25$\pm$0.12)$\times10^{41}$ & (6.93$\pm$0.70)$\times10^{40}$ & (2.67$\pm$0.26)$\times10^{40}$ \\
\noalign{\smallskip}
   $ M_{\rm gas} $ [M$_{\odot}$]  & (5.1$\pm$0.5)$\times10^{6}$ & (1.7$\pm$0.2)$\times10^{6}$ & (1.2$\pm$0.1)$\times10^{4}$ & (5.7$\pm$0.5)$\times10^{2}$\\ 
\hline 
\noalign{\smallskip}
\end{tabular}
\caption{Kinematical and physical properties of the four kinematical components found for the warm ionized gas. The central velocity $v$ and FWHM are obtained from the {\rm[O}\,{\rm \scriptsize III}{\rm]} model, the electron density $n_{\rm e}$ and reddening \textit{E(B-V)} values are extracted using the density diagnostic diagram in Sec.~\ref{DDD_section}, $L(H\beta)$ is the reddening corrected H$\beta$ luminosity, and $M_{\rm gas}$ is the mass of warm ionized gas estimated in Sec.~\ref{The warm ionized gas} .}                                     

\label{Table0}
\end{table*}

The host galaxy profile was then corrected for the differences in the widths of the slices and was subtracted from the ionized gas profiles. The residuals were then fitted with a Gaussian function. For both these profiles we obtained an FWHM of 0.95$\pm$0.01 arcsec, which is consistent with the FWHM of the seeing (i.e., 0.97$\pm0.06$ arcsec, see Sec.~\ref{DataReduction}). This indicates that the warm ionized gas (of all the kinematical components) is not spatially resolved by our observations using this technique and is concentrated in the nuclear regions of the host galaxy.

To obtain an upper limit on the radius of the warm ionized gas, we used the following equation:
\begin{equation}
r \leq \frac{1}{2} \sqrt{({\rm FWHM}+3\sigma)^2-{\rm FWHM}^2}
,\end{equation}  
where the FWHM is the seeing and $\sigma$ is the uncertainty on the seeing. We obtained $r\leq0.3~\rm{arcsec,}$ which at the redshift of the galaxy is equivalent to $r\leq\rm{955~pc}$.

\section{Gas kinematics in the inner regions}\label{The properties of the central BH}

Using the spectro-astrometry technique, we studied the spatial extents of the different ionized gas components, overcoming the limitations given by the seeing. Spectro-astrometry uses high S/N long-slit spectra to measure the centroid position of an unresolved object as function of wavelength. It is based on the fact that the centroid position can be measured with much higher precision than the seeing-limited spatial resolution of the observations \citep{1998SPIE.3355..932B}, and has been used to identify and study binary stars \citep[see][]{1998MNRAS.301..161B,2003A&A...397..675T}. 

We used this technique on the high S/N \OIII$\lambdaup$5007\AA\ line in our slit spectrum and investigated how the warm ionized gas at different velocities is distributed along the slit in the spatial direction. In this way, we probed the gas distribution at subarcsecond scales, which at the redshift of PKS~B1934-63 correspond to scales of tens of parsec.
In the slit spectrum, we isolated the region around the \OIII$\lambdaup$5007\AA\ line, and for a given pixel along the spectral direction, we extracted a profile of the ionized gas along the spatial direction. Every spatial profile probes warm ionized gas at a different velocity. Then, we fit each extracted spatial profile with a Gaussian function and used the fitted profile center to establish the spatial location of the gas at that specific velocity.

\begin{figure}[]
\centering
\includegraphics[width=\hsize, keepaspectratio]{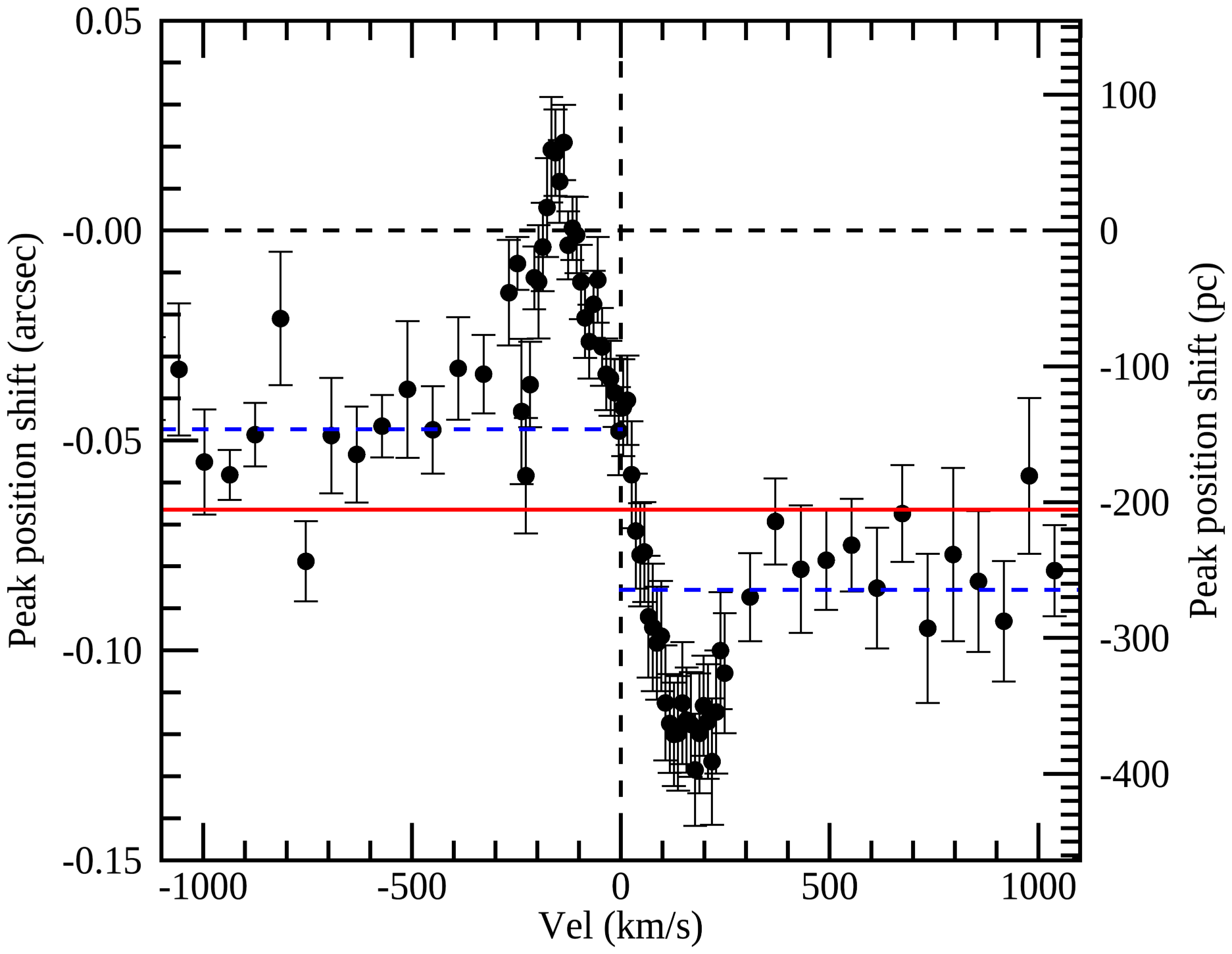}
\caption{Fitted centers of the {\rm[O}\,{\rm \scriptsize III}{\rm]}5007\AA\ emission line spatial profiles, measured in arcsec (left y-axis) and parsec (right y-axis), as a function of velocity. The black dashed vertical and horizontal lines mark the zero point of both axes. The blue dashed lines mark the error-weighted mean position of the \OIII\ spatial profiles at v$<250$\kms\ and at v$>250$\kms\ , and the red solid line marks their average. The negative values along the y-axis indicate the direction pointing toward the companion galaxy to the west.} 
\label{spectroastro}
\end{figure}

We used an average spatial profile of the galaxy starlight to locate the host galaxy center, and we took this as a reference point to establish the location of the ionized gas. We also subtracted the starlight spatial profile from the ionized gas spatial profiles to avoid contamination from the light of the host galaxy. This was particularly important for the profiles extracted at velocities $v<-250$~\kms and $v>$250~\kms , where the fainter \OIII$\lambdaup$5007\AA\ emission of the broad (i.e., the intermediate and very broad) components of the warm ionized gas is located.
To increase the S/N of the gas spatial profiles at these velocities, we binned the data along the spectral direction using a box that was three pixels wide. For gas at velocities $-250<v<250$~\kms , we instead extracted a spatial profile for every pixel along the spectral direction.

In Fig.~\ref{spectroastro} we show the spatial position of the fitted centers of the gas spatial profiles (expressed in arcsec/pc) as a function of the velocity associated with each profile. The zero point along the x-axis is the systemic velocity of the galaxy, while the zero point along the y-axis is the fitted center of the galaxy spatial profile.

The S-shaped trend we see in Fig.~\ref{spectroastro} for the two narrow components (i.e., at $-250<v<$250~\kms) might be explained by a disk-like structure or a biconical outflow in the central regions of the galaxy (up to about $\pm$200~pc).
The curve is symmetric around the zero velocity value and reaches a maximum spatial shift at about $\pm$150~\kms . The overall curve is clearly spatially shifted with respect to the zero point along the y-axis (i.e., the center of the galaxy), and this might be explained by the effect of obscuring dust, which can potentially shift the position of the galaxy spatial profile peak that we used as an indicator of the true AGN nucleus position.

Unlike \cite{2016MNRAS.459.4259R}, we do not find evidence for gas emission on large scales ($>$1~kpc). However, if the inner structure we detect from spectro-astrometry is a circum-nuclear disk (CND), it would rotate in the same sense as the more extended gas disk seen by \cite{2016MNRAS.459.4259R}, with the western gas rotating away from the observer. Our slit is not aligned with the major axis of the structure that \cite{2016MNRAS.459.4259R} observed, which may explain why this is, instead, spatially unresolved by our observations, as shown in Sec.\ref{The radius of the narrow and broad gas components}. 
The CND of warm and cold gas, extended on scales of a few hundreds of parsec, at the center of AGN has previously been found in studies of, for example, \cite{Dumas2007}; \cite{Hicks2013}; \cite{Maccagni2016} and \cite{GarciaB2016}, and it has been proposed that these structures constitute the reservoir of gas from which the BH feeds.
It is worth noting that the amplitude of the rotation that we find is larger than reported in \cite{2016MNRAS.459.4259R}. The reason is probably that their observations do not resolve the double peak of the \OIII$\lambdaup$5007\AA\ line and they used a single Gaussian fit to derive the \OIII$\lambdaup$5007\AA\ velocity field, which smoothes out the velocity gradient that we observe. Considering this, it is plausible that the CND we observe, and the large-scale disk found by \cite{2016MNRAS.459.4259R}, are part of the same disk-like structure whose kinematical axis is misaligned with respect to the axis of the radio source \citep[see Fig.~31 in][]{2016MNRAS.459.4259R}.

On the other hand, the gas at $v<-$250~\kms\ and at $v>$250~\kms\ is associated with the intermediate and the very broad components and is representative of the warm ionized gas that is outflowing.
The error-weighted mean positions of the gas at $v<-$250~\kms\ and at $v>$250~\kms\ are $-$0.047$\pm$0.002 and $-$0.085 $\pm$0.003, respectively (see Fig.~\ref{spectroastro}). This spatial asymmetry supports the idea that the geometry of the outflowing gas is biconical.  
Assuming a biconical geometry (i.e., the blueshifted and redshifted gas emission comes from the two sides of the nucleus), we can estimate the position of the nucleus (i.e., the red horizontal line in Fig.~\ref{spectroastro}) and form an idea of the radius of the outflowing gas. In this way, we find that the outflow has a radius of 59$\pm$12 pc. This matches the radial extent of the radio source well (the separation between the radio lobes of PKS~B1934-63 is 131,7$\pm$0.9~pc) and suggests that the warm ionized gas is outflowing as a consequence of the interaction with the radio jets. Given the orientation of the radio source, it is likely that the jets are accelerating part of the gas in regular rotation in the CND. Together with the findings of \cite{2016MNRAS.459.4259R}, this suggests that the velocity gradient of the kinematically disturbed gas is aligned along the radio axis.

We are aware that taking only the average position for the gas in the broad wings of the \OIII$\lambdaup$5007\AA\ line might underestimate the size of the outflow. However, even considering the higher absolute value for the shift of the gas at $v<-$200~\kms\ and at $v>$200~\kms\ with respect to the estimated position of the nucleus, we find that if we were to include the errors on the spatial shifts, the outflow would have a maximum radius of $\sim 175$~pc. This indicates that the outflow is extended on spatial scales that are comparable with the size of the radio source.

We can exclude that the motions of the gas in the broad wings of the \OIII$\lambdaup$5007\AA\ line are due to rotation around the central BH. If the gas at velocities between 500 and 1000 \kms\ is located at a distance of about 60~pc from the central BH, a BH mass in the interval 0.34-1.35$\times$10$^{10}$~M$_{\odot}$ would be required. Such masses are too high for the BH of a galaxy with a stellar mass of about 10$^{11}$~M$_{\odot}$ \citep[as estimated by][]{2016MNRAS.459.4259R}, which according to scaling relations \citep{Kormendy2013}, is expected to host a BH with a mass of about 5$\times$10$^{8}$~M$_{\odot}$.

As a sanity check, we also performed a spectro-astrometry analysis on the \OIII$\lambdaup$4958\AA\ line. The gas at $-$250$<v<$250~\kms confirms the trend seen for the \OIII$\lambdaup$5007\AA\ line. However, although the lower S/N of the broad wings of the line does not allow us to confirm the exact spatial extent of the outflow component, in line with our findings for the \OIII$\lambdaup$5007\AA\ line, we can see that this gas is extended on smaller spatial scales than the gas at low velocities.

\section{Warm ionized gas and outflow parameters}\label{The warm ionized gas}

We estimated the warm ionized gas mass of the different kinematical components using the following equation:

\begin{equation}
M_{\rm gas}= \frac{L({\rm H}\beta)m_{\rm p}}{n_{\rm e} \alpha^{\rm eff}_{\rm H\beta} h \nu_{\rm H\beta}}
,\end{equation}
where $L({\rm H\beta})$ is the $\rm{H\beta}$ luminosity corrected for dust extinction, $m_{\rm p}$ is the proton mass, $n_{\rm e}$ is the electron density from the density diagnostic diagram, $\alpha^{\rm eff}_{\rm H\beta}$ is the effective $\rm{H\beta}$ recombination coefficient \citep[taken as 3.03$\times10^{-14}~\rm{cm^3 s^{-1}}$ for case B in the low-density limit;][]{2006agna.book.....O}, $\nu_{\rm H\beta}$ is the frequency of the $\rm{H\beta}$, and $h$ is the Planck constant.
The $L({\rm H\beta})$ and the estimated $M_{\rm gas}$ of each kinematical component are reported in Table \ref{Table0}.

We found that the two narrow components have a mass of warm ionized gas of $M_{\rm gas}($1N$)=(5.1\pm 0.5)\times 10^{6}$~M$_{\odot}$ and $M_{\rm gas}($2N$)=(1.7\pm 0.2)\times10^{6}$~M$_{\odot}$. The intermediate component has a gas mass of $M_{\rm gas}($I$)=(1.2\pm 0.1)\times10^{4}$~M$_{\odot}$, while for the very broad component, we found $M_{\rm gas}($VB$)=(5.7\pm 0.5)\times10^{2}$~M$_{\odot}$.
It is clear that almost the entire reservoir of the host galaxy's warm ionized gas is found in the two narrow components; the intermediate and the very broad components represent only a small fraction of the warm ionized gas reservoir. 

With a reliable estimate of the electron density of the outflowing warm ionized gas, we were able to characterize the properties of the outflow.
Following the method described in Sec.~4.1 of \cite{Rose2017}, we determined the mass outflow rate $\dot{M}$, the outflow kinetic power $\dot{E}$ and the AGN feedback efficiency $F_{\rm kin}=\dot{E}/L_{\rm bol}$ using the following formulae:

\begin{equation}
\dot{M}= \frac{ L({\rm H}\beta)m_{\rm p} v_{\rm out}}{n_{\rm e} \alpha^{\rm eff}_{{\rm H}\beta} h \nu_{{\rm H}\beta} r }
\label{Mdot}
\end{equation}  

\begin{equation}
\dot{E}=\frac{\dot{M}}{2} (v^2_{\rm out}+ 3\sigma^2)
\label{Edot}  
,\end{equation}

where $v_{out}$ is the velocity, $r$ is the radius, and $\sigma$ is the line-of-sight velocity dispersion ($\sigma={\rm FWHM}/2.355$) of the outflow.
  
To estimate these parameters, we used the radii estimated in Sec.~\ref{The radius of the narrow and broad gas components} (i.e. $r\rm{<955~pc}$) and in Sec.\ref{The properties of the central BH} (i.e., $r\rm{>60~pc}$) as upper and lower limits for the radius of the outflow, respectively.
We extracted the bolometric luminosity using the dereddened \OIII$\lambdaup$5007\AA\ total luminosity, which is usually considered a good indicator of the AGN power \citep[see][]{2004ApJ...613..109H}. Summing the intrinsic  \OIII$\lambdaup$5007\AA\ luminosities of each component, we obtained ${L_{\OIII}=(4.4\pm 0.4)\times10^{42}}$~\ergs.
To extract the bolometric luminosity, we used the bolometric correction of \cite{2009A&A...504...73L}, $L_{\rm bol}=454~L_{\OIII}$ , which is valid for object with intrinsic \OIII$\lambdaup$5007\AA\ luminosity in the interval $L_{\OIII}=10^{42 - 44}$~\ergs. This resulted in a bolometric luminosity of $L_{\rm bol} = (2\pm 0.2)\times 10^{45}$~\ergs\ , which makes PKS~B1934-63 a type II quasar according to the criterion of \cite{2003AJ....126.2125Z}.

We estimated the outflow properties for the gas emitting the very broad component and for both of the broader components together (i.e., the intermediate and very broad component). In the latter case, for $L(H\beta)$ we took the intrinsic integrated $H\beta$ luminosity of the two components, while for $n_{\rm e}$, $v_{\rm out}$ , and the FWHM of the gas, we took a flux-weighted value. All the relevant quantities for these calculations are reported in Table~\ref{Table1_2} together with the estimated $\dot{M}$, $\dot{E}$, and $F_{\rm kin}$. In this way, we obtained mass outflow rates in the range 10$^{-3}$-10$^{-4}~\rm{M_{\odot}~yr^{-1}}$ and AGN feedback efficiencies in the range 10$^{-4}$-10$^{-5}\%$. 

This approach assumed that the central velocity of the line associated with the outflowing gas is the true velocity of the outflow $v_{\rm out}$ and its broadening is due to the gas turbulence. A less conservative approach, which can potentially take into account projection effects, considers that the broadening of the lines is due to the different projections of the velocity vectors of the gas in a quasi-spherical outflow.
In this case, the actual outflow velocity $v_{\rm out}$ is the maximum velocity that the gas reaches in the wings of the emission line profile. To estimate the maximum velocity of the gas, we followed the approach of \cite{Rose2017} and took the velocity corresponding to a 5$\%$ cut of the flux in the blueshifted direction.
We report all the relevant quantities for the calculations for this case as well, together with the estimated $\dot{M}$, $\dot{E}$, and $F_{\rm kin}$ in Table~\ref{Table1_2}.
With this method we obtained mass outflow rates and AGN feedback efficiencies for the intermediate and very broad components, which are, on average, one order of magnitude higher than the results obtained with the first approach.

Both methods gave AGN feedback efficiencies that are among the lowest found for warm ionized gas outflows \citep[see Fig.2 in][]{Harrison2018}. Our values are far from the 5-10$\%$ required by the classical AGN feedback models \citep[e.g.,][]{1999MNRAS.308L..39F,2005Natur.433..604D,2005MNRAS.361..776S} and also lower that the 0.5$\%$ of the multi-staged model proposed by \cite{2010MNRAS.401....7H}. 
However, as stressed by \cite{Harrison2018}, there are some caveats to consider when comparing the AGN feedback efficiency derived from observations to the prescription of cosmological models, especially because the energy that is actually transferred to the warm and cool ISM can be a fraction of the energy injected by the AGN into the surrounding medium.

\begin{table*} 
\renewcommand{\arraystretch}{1.5}
\centering 
\begin{tabular} { p{2.7cm}  |  P{3.4cm} P{4cm}  |  P{3.5cm} P{3cm} } 

\hline              
        \textbf{ }                                                                                                       & \textbf{VB Component}                          & \textbf{I+VB Component}  & \textbf{VB Component}  & \textbf{I Component} \\    
                \textbf{ }                                                                                                       &                        &   & v$_{\rm max}$ & v$_{\rm max}$ \\ 
\hline\hline
        \textbf{$L({\rm H\beta})~[\rm{erg~s^{-1}]}$}                                            & $2.6\times10^{40} $ & $ 9.6 \times10^{40}$      & $2.6\times10^{40}$ & $6.9\times10^{40}$ \\  

    $v_{\rm out}$~[\kms ]    &          302     &       156      & 2462 & 777  \\  
 
   $n_{\rm e} ~\rm{[cm^{-3}]}$          &   $1\times 10^{5.5}$  &  $1.7\times 10^{5}$ & $1\times 10^{5.5}$ & $1\times 10^{4.6}$  \\              
   FWHM~[\kms ]                                         &  2035 &       568  & -- & -- \\
\hline  
   $\dot{M}~{\rm [M_{\odot} yr^{-1}]}$          & $0.002$-$0.0002$    &  $0.009$-$0.0006$   & $0.0015$-$0.02$  & $0.01$-$0.14$ \\
 
   $\dot{E}~\rm{[erg~s^{-1}]}$                  & $     (0.13$-$2.0)\times 10^{39} $  & $(0.2$-$6.7) \times 10^{39}$       &   $(0.29$-$4.2)\times 10^{40}$ &  $(0.18$-$2.7)\times 10^{40}$  \\
    
   $F_{\rm kin}$                         & $(0.07$-$1)\times10^{-6}$  &          $ (0.1$-$3.3) \times10^{-6}$ & $ (0.14$-$2.11) \times10^{-5}$  &  $ (0.1$-$1.3)\times10^{-5}$ \\   

\hline  
\end{tabular}
\caption{Mass outflow rates $\dot{M}$, outflow kinetic energy $\dot{E,}$ and AGN feedback efficiency $F_{\rm kin}=\dot{E}/L_{\rm bol}$ obtained with the two different methods described in the text, together with all the relevant quantities used in the calculation. Columns 2 and 3 report the values obtained using the central velocity of the outflowing gas as $v_{\rm out}$, while Cols. 4 and 5 report the values obtained using the maximum velocity $v_{\rm max}$ as $v_{\rm out}$.}
\label{Table1_2}
\end{table*}

\section{Gas ionization}\label{ionization_section}

Given that the gas in the intermediate and the very broad components shows remarkable differences in terms of densities and kinematics from the rest of the warm ionized gas, we investigated whether these components also stand out in terms of their ionization properties.

\begin{figure*}[t]
\centering
\includegraphics[width=\hsize]{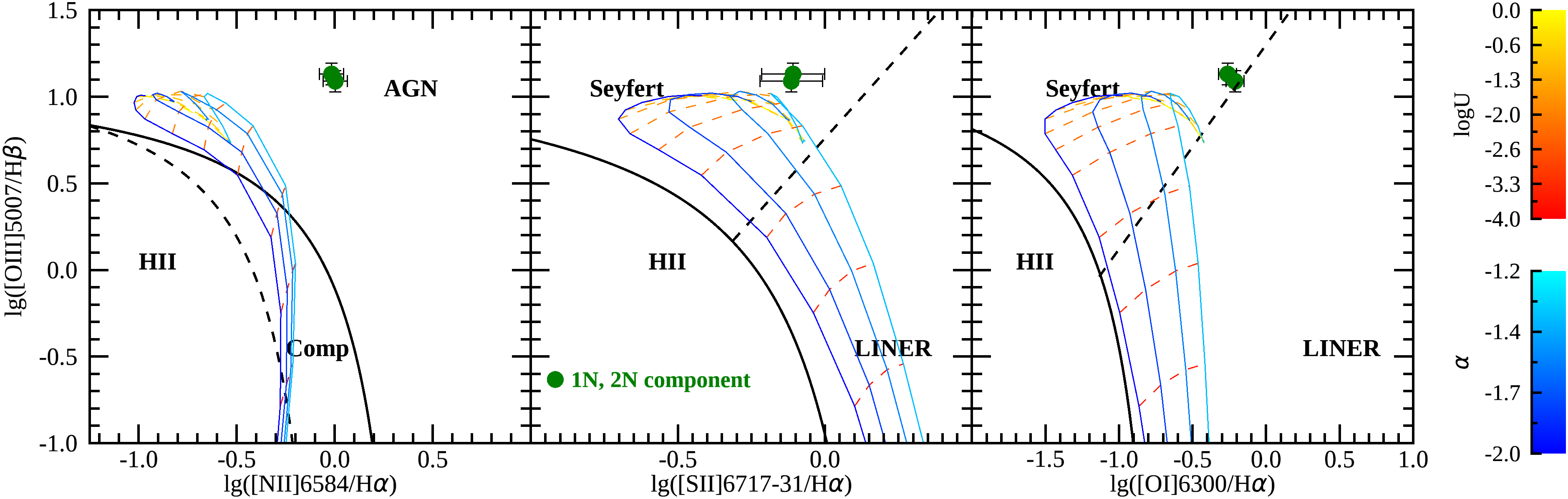}

\includegraphics[width=\hsize]{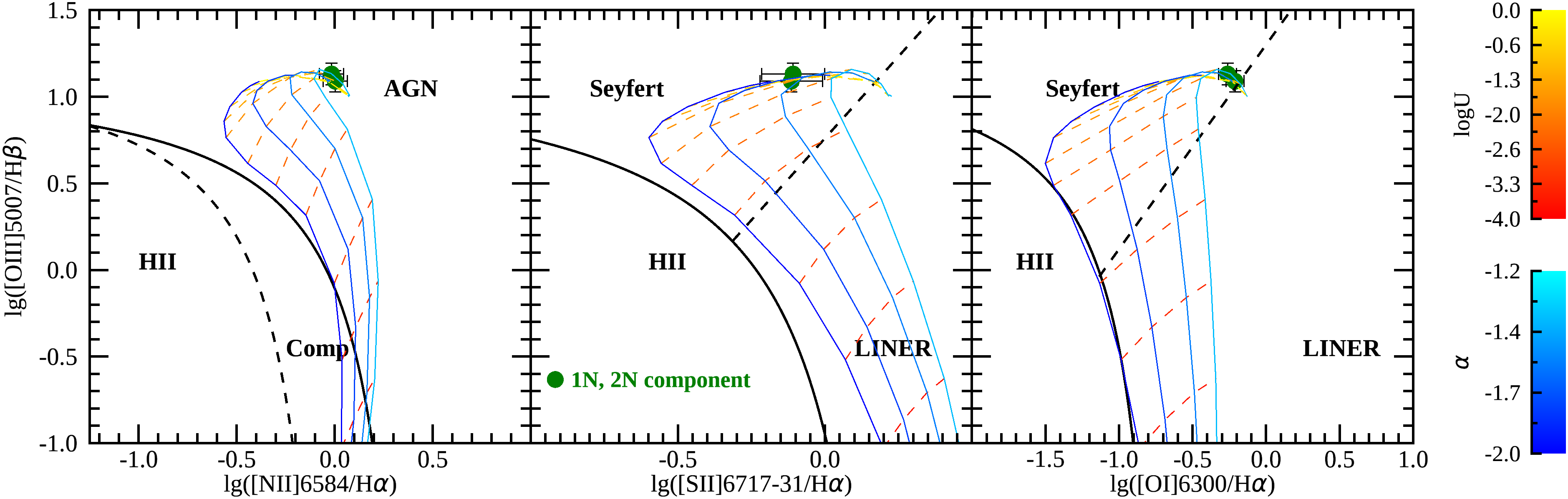}

\caption{BPT diagrams for the two narrow components with models of AGN photoionization with solar metallicity (\textit{upper panels}) and twice the solar metallicity (\textit{lower panels}).  
The models have a gas density $n_{\rm e}=10^3\rm{cm^{-3}}$, the dashed lines indicate models with constant photoionization parameter ${\rm log}U$ (from $-4$ to 0, from bottom to top ), and the solid lines refer to models with constant spectral index $\alpha$ (from $-$2 to $-$1.2, from left to right).
The solid line in all the panels is the \cite{2001ApJS..132...37K} maximum starburst line. The dashed line in the left panels is the semi-empirical  \cite{2003MNRAS.346.1055K} line. The dashed line in the central and right panels is the empirical \cite{2006MNRAS.372..961K} line separating Seyfert galaxies from LINERS.}
\label{class_BPT}
\end{figure*}

\begin{figure*}[t]
\centering
\includegraphics[width=\hsize]{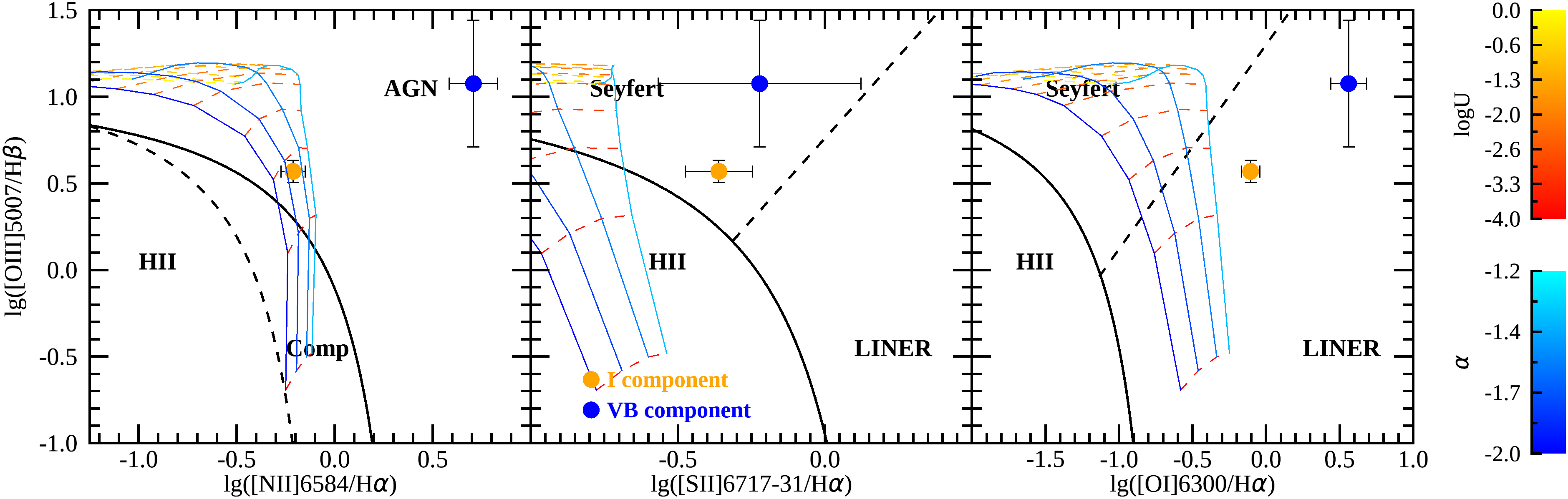}

\includegraphics[width=\hsize]{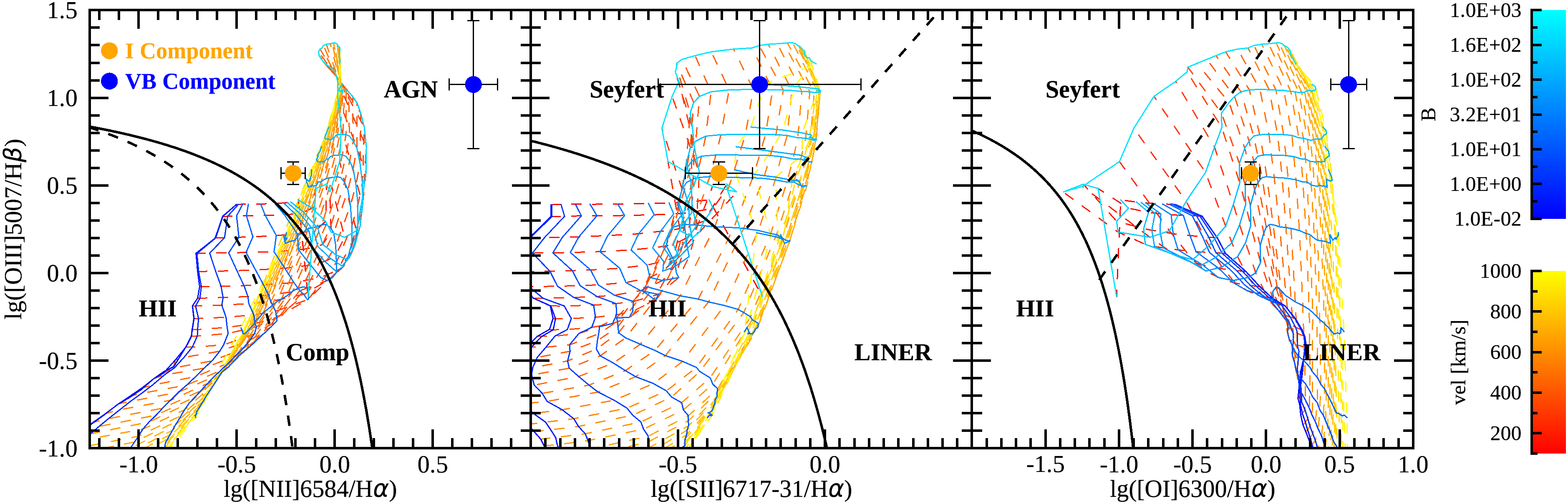}
\caption{BPT diagrams for the intermediate and very broad components with models of AGN photoionization (\textit{upper panels}) and shocks (\textit{lower panels}) with solar metallicity.  
The AGN photionization models have a gas density $n_{\rm e}=10^5\rm{cm^{-3}}$, and the $\alpha$ and ${\rm log} U$ parameters vary in the same range as described in Fig.~\ref{class_BPT}.
The shock models have a pre-shock gas density of $n_{\rm e}=10^3\rm{cm^{-3}}$, the dashed lines indicate models with constant shock velocity $v_{\rm s}$ (ranging in the interval 100-1000~\kms, from left to right), and the solid lines refer to models with a constant magnetic parameter $B$ (ranging in the interval 0.01-1000 $\upmu$G, from bottom to top). The solid and dashed black lines are the same as in Fig.~\ref{class_BPT}. }

\label{first_shock}
\end{figure*}

In Figs.~\ref{class_BPT}  and~\ref{first_shock} we present the classical BPT diagrams originally introduced by \cite{1981PASP...93....5B} with the observed line ratios of the different gas kinematical components.
As expected, the observed line ratios of the four kinematical components are typical of AGN, although with some differences between the components. As in the case of the density diagnostic diagram, the two narrow components have similar line ratios (see Fig.~\ref{class_BPT}), confirming that they are part of the same structure.

All the gas components, but in particular the very broad component, have high values of the \OI$\lambdaup$6300/H$\alpha$ line ratio, higher than those usually measured in AGN, and they are indicative of shock-ionized gas. The \OI$\lambdaup$6300\AA\ line is emitted by warm weakly ionized gas that is typically located in the transition region between ionized gas and neutral gas. Only high-energy photons can penetrate this region, which stimulates the \OI$\lambdaup$6300\AA\ emission, and a source of such high-energy photons can be the AGN continuum radiation, shocks, or a combination of the two.

We investigated the gas ionization mechanisms by comparing the observed line ratios with model grids of AGN photoionization and of shocks. The model grids were taken from the ITERA tool \citep{2010NewA...15..614G} and created using {\rm MAPPINGS III} \citep{2013ascl.soft06008S}. 
The AGN photoionization model grids (see Fig.~\ref{class_BPT} and the upper panel in Fig.~\ref{first_shock}) were obtained by varying the spectral index $\alpha$ from $-2$ to $-1.2$  and the ionization parameter $ \log U$ from $-4$ to 0. 
The shock models (see the lower panel in Fig.~\ref{first_shock}) included in the ITERA tool were taken from \cite{2008ApJS..178...20A}, they have solar metallicity and cover shock velocities in the range $v_{\rm s}$=100-1000~\kms\ and magnetic fields (i.e., pre-shock transverse magnetic field) in the range $B$=0.01-1000 $\upmu$G.

In Fig.~\ref{class_BPT} we compare the line ratios of the narrow components to AGN photoionization models with gas density $n_{\rm e}=10^3~\rm{cm^{-3}}$ (according to the densities derived in Sec.~\ref{DDD_section}) and both solar and twice solar (i.e., 2~$Z_{\odot}$) metallicity. 
The comparison  of the line ratios with the photoionization models provides good evidence for super-solar metallicities in the near-nuclear regions. We also found a good consistency between the positions of the points and the models (in $U$ and $\alpha$) in all three plots for twice solar metallicity.
The narrow components do not show kinematical evidence of shocked gas (i.e., broad line width), and for this reason, we did not compare their line ratios with shock models.

It is possible that the gas of the narrow components is part of the shock precursor. However, for solar metallicities, precursor models have line ratios similar to the AGN photoionization models and would also fail to explain the observed line ratios. We cannot comment on precursor models with metallicities higher than solar because there are no such models.

In Fig.~\ref{first_shock} we use both AGN photoionization and shock models to study the ionization state of the gas in the intermediate and the very broad components.
We considered AGN photoionzation models with a gas electron density $n_{\rm e}=10^5~\rm{cm^{-3}}$, this model grid was not included in the ITERA tool and was produced using Cloudy \citep[C13.04, ][]{2013RMxAA..49..137F} photoionization code. For the shock models we assumed a shock compression factor of 100 and took a pre-shock gas electron density of $n_{\rm e}=10^{3}~\rm{cm^{-3}}$. The assumed compression factor was intended to take into account the fact that while shock conditions cause a modest density jump (maximum factor $\sim$5, see Fig.~7 in \citealp{2017ApJS..229...34S}), the temperature jump will further increase the pressure and thus the compression of the gas in the post-shock regions.
We found that overall, the shock models reproduce the observed line ratios in the three BPT diagrams (see Fig.~\ref{first_shock}) better than the AGN photoionization models.
In particular, the \SII$\lambdaup\lambdaup$6717,31/H$\alpha$ and the \OI$\lambdaup$6300/H$\alpha$ line ratios of the intermediate component can be explained by shock models with velocities $v_{\rm s}$=400-500~\kms .

The line ratios of the very broad component were more difficult to explain with current models.
The outflowing gas of the very broad component stands out from the rest of the warm ionized gas in terms of both the \NII$\lambdaup$6584/H$\alpha$ and the \OI$\lambdaup$6300/H$\alpha$ line ratios, but not in terms of the \SII$\lambdaup\lambdaup$6717,31/H$\alpha$ ratio.
This can be explained with the high densities that are associated with this gas component and with the fact that the \SII\ lines have a lower critical density \citep[about 5$\times 10^3\rm{cm^{-3}}$,][]{1988ApL&C..27..275Z} than the \NII\ and \OI\ emission lines.
Considering the trend of the model grids, the \OI$\lambdaup$6300/H$\alpha$ line ratio is consistent with shocks of higher velocities ($v_s\geq$1000~\kms), as is also suggested by the extreme kinematics of the gas of the very broad component. On the other hand, the high value of the \NII$\lambdaup$6584/H$\alpha$ line ratio might be an indication of and enhanced N/O ratio (i.e., higher than solar) for the gas of the very broad component \citep[see, e.g.,][]{1994MNRAS.268..989T,2017arXiv170907000M}.

We are aware that the variation in the shock compression factor and gas metallicity plays a role in the final line ratios of the models. In addition, degeneracies in the fitting of the intermediate and very broad components in the \NII\ + H$\alpha$ blend may contribute to the extreme \OI /H$\alpha$ and \NII /H$\alpha$ ratios that are observed.
New shock models with a fully self-consistent treatment of the pre-shock ionization and thermal structure are being developed for fast shocks and high gas electron densities \citep{2017ApJS..229...34S}.

The \OIII$\lambdaup\lambdaup$4958,5007/4363 ratio, classically used as a temperature diagnostic, together with the \HeII$\lambdaup$4686/H$\beta$ ratio give us further indications that the temperature of the gas of the intermediate component is high and that this is due to shocks. Unfortunately, we cannot comment on the temperatures associated with the gas of the very broad component because the detection of this component in the \OIII$\lambdaup$4363\AA\ line depends on the model used to fit the H$\gamma$ line. When we add a broad redshifted component for H$\gamma$ (according to what we find for the H$\beta$ line) to the fitting procedure, the very broad component of the \OIII$\lambdaup$4363\AA\ line is not detected (see Fig.~\ref{Hg_OIII}). In addition, the very broad component is not detected in the case of the \HeII$\lambdaup$4686\AA\ emission line (see Fig.~\ref{HeII}).

By taking into account the main error sources in the fluxes of the \OIII$\lambdaup$4363\AA\ components (i.e., the model used for the fitting of the H$\gamma$ line and the level of the line continuum), we obtain an \OIII$\lambdaup\lambdaup$4958,5007/4363 ratio of 70$\pm$2 and 104$\pm$3 for the two narrow components and of 30$\pm$3.6 for the intermediate component. 
To derive an electron temperature for the gas of the narrow components and the gas of the intermediate component, we used the formula given in \cite{2006agna.book.....O} and an electron density of $n_e$=10$^{3}$~cm$^{-3}$ and $n_e$=10$^{4.5}$~cm$^{-3}$ , respectively, according to our findings in Sec.\ref{DDD_section}.
We find that these line ratios correspond to gas electron temperatures of $T_{e}(1N)$=15100$\pm$200~K, $T_{e}(2N)$=13000$\pm$200~K, and $T_{e}(I)$=30175$\pm$4400~K.

As shown by the models of \cite{Binette1996}, it is possible to have high electron temperatures in the presence of matter-bounded clouds that are photoionized by the AGN continuum. However, as stressed in \cite{VillarMartin1999}, these models cannot explain temperatures higher than 20000~K \textit{\textup{and}} a \HeII$\lambdaup$4686/H$\beta$ ratios lower than about 0.3-0.4 simultaneously.
For the intermediate component we find a \HeII$\lambdaup$4686/H$\beta$=0.07$\pm$0.04, which together with the high temperature favors shocks as the mechanism that heats the gas, in agreement with our findings from the BPT diagram shown in Fig.~\ref{first_shock}.

Additional evidence for shocks is also provided by the \FeII$\rm{\lambdaup}$1.257$\upmu$m/Pa$\beta$ line ratio. This line ratio can indicate whether shock ionization produces the \FeII\ emission.
This was first suggested by \cite{1993ApJ...416..150F} and \cite{1994ApJ...421...92B}, who found a correlation between the \FeII\ and the radio emission in radio AGN. It has been shown that in galaxies hosting an AGN, a  \FeII$\rm{\lambdaup}$1.257$\upmu$m/Pa$\beta$ ratio close to 0.6 is produced by AGN photoionization, while a ratio close to 2 is related to shock excitation \citep{1999MNRAS.304...35S,2004A&A...425..457R}.

In the NIR band of the nuclear spectrum of PKS~B1934-63, we detected both the \FeII$\rm{\lambdaup}$1.257~$\upmu$m and Pa$\beta$ lines. Because their S/N is low, we were unable to fit the lines with the \OIII\ model (see Fig.~\ref{pabeta_ironlines}), and we extracted a line ratio using the total flux of the two lines. We find \FeII 1.257/Pa$\beta$=1.44 $\pm$0.2, which is indicative of shock-ionized gas. 

Overall, the combination of the diagnostic diagrams presented in Fig.~\ref{first_shock} and the \OIII$\lambdaup\lambdaup$4958,5007/4363, \HeII$\lambdaup$4686/H$\beta$ and \FeII$\rm{\lambdaup}$1.257$\upmu$m/Pa$\beta$ ratios strongly support the idea that the intermediate outflow component is shock ionized.

\section{H$_{2}$ warm molecular and the neutral gas}\label{The H2 warm molecular and neutral gas}

\begin{figure}[]
\centering
\includegraphics[width=\hsize, keepaspectratio]{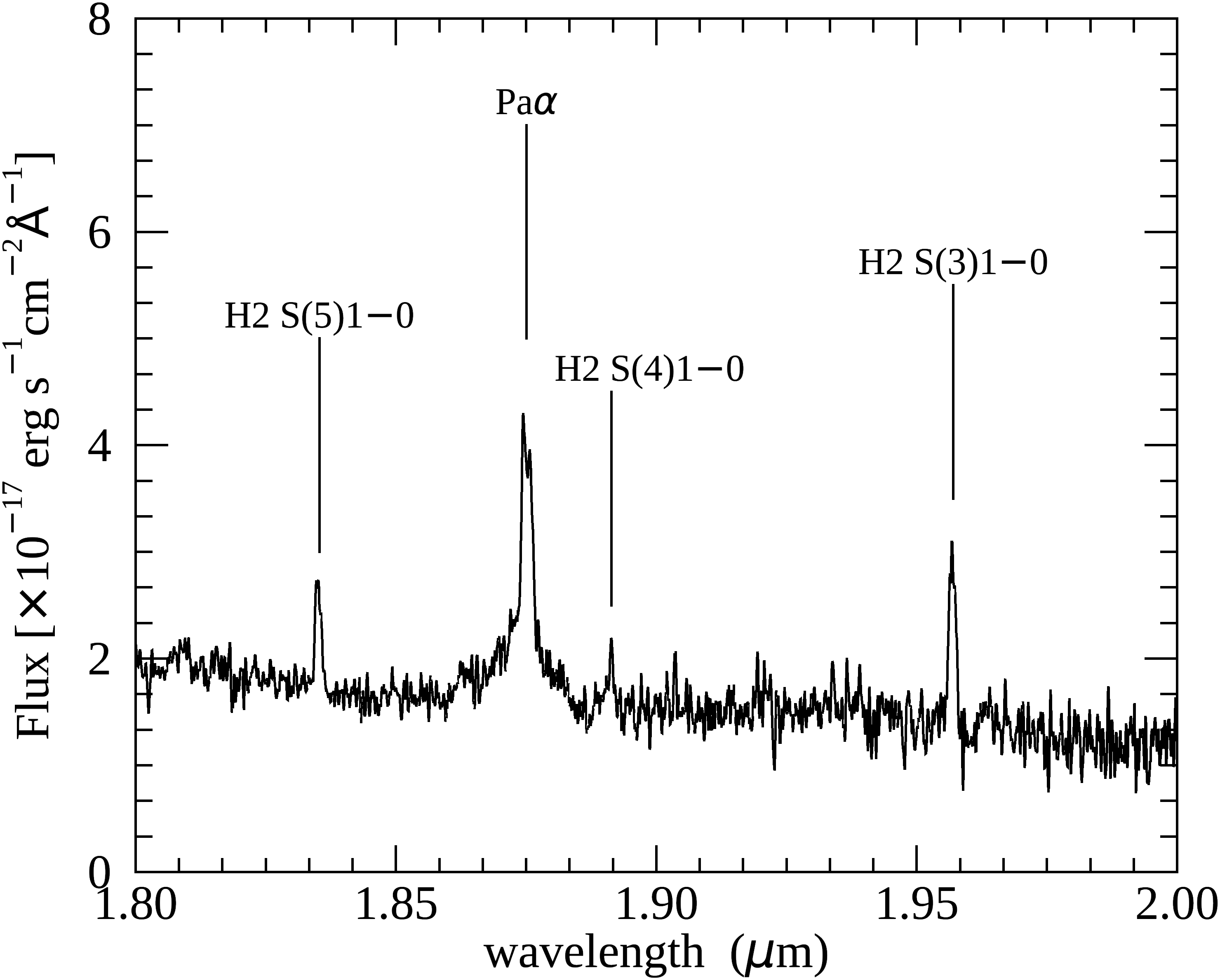}
\caption{Section of the nuclear spectrum of PKS1934-63 in the NIR band showing the warm molecular \Hdue\ emission lines and the Pa$\alpha$ line. The wavelength and the name of each line are indicated.}
\label{IR_spectrum}
\end{figure}

Compact steep-spectrum radio sources are known to host massive outflows of molecular and atomic gas \citep[see, e.g.,][]{2005A&A...439..521M,2012A&A...541L...7D,2014Natur.511..440T}. Probing only the warm ionized gas phase, we might be missing part of the gas that is outflowing. 
The broad wavelength range covered by X-shooter enabled us to probe the kinematics of the warm molecular gas via the \Hdue\ emission lines detected in the NIR band. In the nuclear spectrum of PKS~B1934-63, we identified the \Hdue\ S(5)1-0 line at 1.835~$\upmu$m, the \Hdue\ S(4)1-0 line at 1.891~$\upmu$m, and the \Hdue\ S(3)1-0 line at 1.957~$\upmu$m. Fig.~\ref{IR_spectrum} clearly shows that the \Hdue\ emission lines are narrower than the warm ionized gas emission lines (e.g., the Pa$\alpha$ line). 

\begin{figure}[]
\centering
\includegraphics[width=\hsize, keepaspectratio]{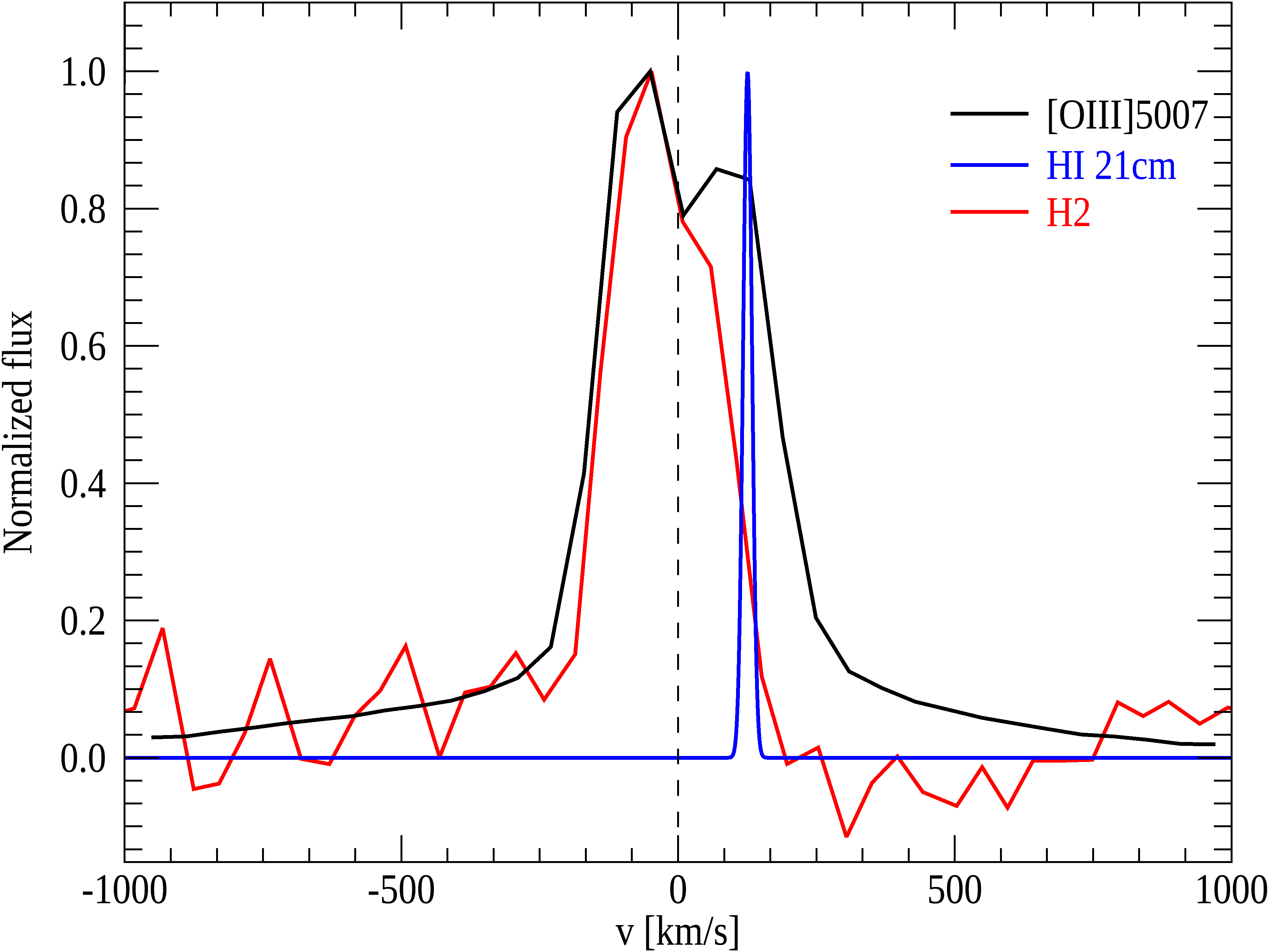}
\caption{Normalized {\rm[O}\,{\rm \scriptsize III}{\rm]}$\rm{\lambdaup}$5007\AA\  line (solid black), stacked \Hdue\ line (solid red), and \HI\ 21~cm line (solid blue) in velocity space. Each line is normalized with respect to its maximum value. The zero velocity along the x-axis is the systemic velocity of the galaxy as extracted in Sec.~\ref{z_stellarpop} and is marked with the black vertical dashed line. The {\rm H}\,{\rm \scriptsize I} line has been observed in absorption by \cite{2000A&A...362..426V} and is reproduced as an emission line using a Gaussian function and the parameters reported in the text.}
\label{velspace}
\end{figure}

To provide an overview of the kinematics of the different phases of the gas, in Fig.~\ref{velspace} we compare the kinematics of the warm ionized gas with the kinematics of the warm molecular and atomic gas by plotting the normalized line profiles of the \OIII$\lambdaup$5007\AA\ , the \Hdue\ , and the \HI\ 21~cm spectral lines together. 
The \Hdue\ profile in Fig.~\ref{velspace} is the stacked profile of the \Hdue\ 1.957~$\upmu$m  and   \Hdue\ 1.835~$\upmu$m lines and  appears to be slightly blueshifted with respect to the systemic velocity of the host galaxy. Its peak coincides with the peak of the blueshifted narrow component of the ionized gas and does not show clear signs of kinematically disturbed gas, but because of the low S/N and the fluctuations of the continuum, we cannot completely rule out that there might be outflowing gas.

\HI\ gas has been detected in absorption against the radio continuum source by \cite{2000A&A...362..426V} and is shown inverted to emission for easy comparison in Fig.~\ref{velspace}. These observations trace only the kinematics of gas that is located, in projection, in front of the compact radio source. The \HI\ has a velocity shift ($v$=116~\kms ), which is comparable to the redshifted narrow component of the warm ionized gas and is characterized by a very narrow profile (FWHM=18.8~\kms ). This indicates that it might be connected to infalling clouds of atomic gas located in front of the radio source \citep[as in the case of PKS B1718-649, see][]{2014A&A...571A..67M}.

A more global view on the atomic gas phase of the ISM could be obtained by using the absorption lines of the \NaD\ doublet at $\lambdaup\lambdaup$5890,5896\AA\ \citep[see, e.g.,][]{2011A&A...532L...3L}.
After the stellar continuum of the host galaxy was subtracted, we did not find evidence for the \NaD\ absorption in the nuclear spectrum of PKS~B1934-63. 
This might be due to the compactness of the neutral ISM, which is concentrated in the inner regions of the galaxy (like the warm ionized gas) and does not absorb its diffuse starlight.

In the UVB part of the spectrum, we detected the ISM absorption features of the \MgII$\lambdaup\lambdaup$2795,2802\AA\ doublet. These absorption lines trace gas in a low-ionization state where the \HI\ is the dominant phase, and are superimposed on the \MgII\ emission lines at the same wavelengths because of the AGN.
Even though we fit the \MgII\ lines with a simple model, which might ignore the complex emission line profile underneath the absorption  (see Fig.~\ref{MgIIfit}), we did not observe clear signs of kinematically disturbed gas in absorption. The \MgII\ emission and absorption were fit by single-Gaussian components with an FWHM$\sim$1250\kms\ and $\sim$200\kms\ , respectively, and the absorption component is centered at the systemic redshift of the galaxy (see Appendix \ref{appendix1} for further details).

We detected such deep \MgII\ absorption, which implies a covering factor of the line-emitting gas close to 1, but no clear evidence for ISM absorption of the stellar continuum at the same wavelength (cf. \CaII, \NaD). This provides further evidence that the emission lines are emitted by a region that is compact relative to the stellar body of the galaxy.

We conclude that while we clearly detect an outflow of warm ionized gas, there is no strong evidence of cold outflowing gas traced by the \Hdue , \HI,\ and \MgII\ lines. 
This is in contrast with compact steep-spectrum radio sources such as IC~5063 \citep{2014Natur.511..440T,2015A&A...580A...1M,Oosterloo2017}, PKS~B1345+12 \citep{2013Sci...341.1082M, 2012A&A...541L...7D}, and 3C~305 \citep{2005A&A...439..521M}, in which ionized gas outflows have also been detected. In these sources, the cold molecular (CO) and warm molecular (\Hdue) gas has been found to be the dominant outflowing phase in terms of mass. 
 
In recent years, the scenario in which cold gas is formed in situ, within the material swept away by the AGN, has gained more acceptance and has also been invoked to explain the properties of the multi-phase outflow of compact steep-spectrum radio sources \citep{2014Natur.511..440T,2015A&A...580A...1M}. According to the latter scenario, the AGN at first drives fast shocks into the ISM, ionizing the gas and heating it to high temperatures (higher than 10$^6$~K). The post-shock gas then cools down, accumulating as atomic and eventually cold molecular gas at a later stage. 
This means that outflows of warm ionized gas should be detected in the early phase of an AGN-ISM interaction, and only when sufficient time has elapsed for a substantial amount of gas to cool would we expect to be able to detect the cold molecular counterpart of these outflows (e.g., using CO lines).
Recent simulations confirmed this scenario, finding that molecular gas starts to form around a few $10^5$~yr from the start of the AGN-ISM interaction \citep{2017arXiv170603784R}.

\begin{figure*}[t]
\centering
\includegraphics[width=\hsize, keepaspectratio]{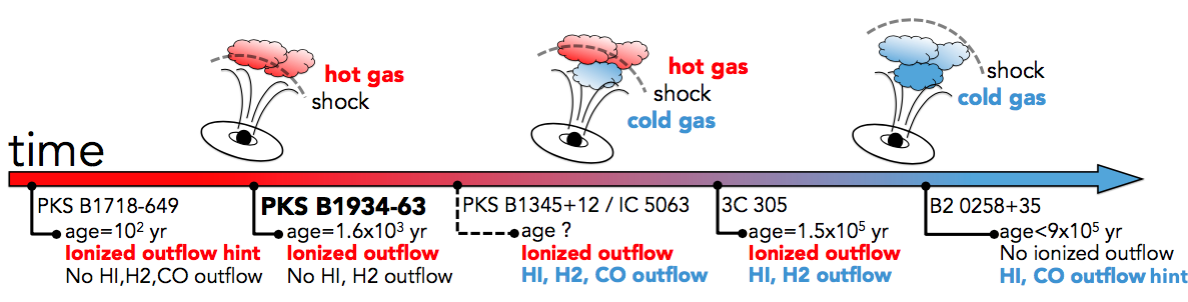}
\caption{Time-line sketch reporting the radio ages and the multi-phase properties of the outflows found in the compact sources PKS~B1718-649,  PKS~B1934-63, PKS~B1345+12, IC~5063, 3C~305, and B2~0258+35 in the context of the evolutionary scenario discussed in Sec.~\ref{The H2 warm molecular and neutral gas}.}
\label{timeline}
\end{figure*}

We performed a first attempt to test this scenario by integrating our findings with the outflow properties reported in the literature for other young compact radio sources. Compact radio sources have radio ages between 10$^2$~yr and 10$^5$~yr \citep{2003PASA...20...19M,2009AN....330..193G}, and we used them to sample outflowing gas at different times. 
When we exclude PKS~B1934-63, our target, only a few other compact young radio sources have both an estimated radio age and observations of their multi-phase ISM. It is worth mentioning that the sources we selected from the literature cover three orders of magnitudes in radio power, and their outflows properties might be due to an intrinsically different type of interaction between the ISM and the radio plasma.

PKS~B1718-649 is a younger \citep[kinematical $\rm{age\sim10^2~yr}$,][]{2009AN....330..193G} compact radio source than PKS~B1934-63, showing hints of warm ionized outflowing gas \citep[blueshifted wings in the forbidden emission lines, see the optical spectrum in][]{1985AJ.....90.1172F}. Multi-wavelength observations find that as for our target, no evidence of outflowing gas in colder phases has been obtained for this source (atomic \HI , \citealp{2014A&A...571A..67M}; warm molecular \Hdue , \citealp{Maccagni2016}; cold molecular CO, \citealp{Maccagni2018}).

An older compact radio source that is able to probe a more evolved stage of the AGN-ISM interaction is 3C~305. It has an estimated radiative age of 1.5$\rm{\times10^5}~yr$ \citep{1999A&A...345..769M} and shows evidence of an outflow in the warm and cold molecular gas \citep[][and Guillard et al in prep.]{2012ApJ...747...95G}, and in the atomic \HI\ and warm ionized gas \citep{2005A&A...439..521M}.

Finally, B2~0258+35 is one of the oldest compact radio sources (radiative age $\leq9\rm{\times10^5}~yr$, \citealp{2005A&A...441...89G}; $\leq5\rm{\times10^5}~yr$ Brienza et al. in prep.). Cold molecular CO and atomic \HI\ gas have been detected \citep{2007ASPC..375..271P,2010A&A...523A..75S} and show signs of disturbed kinematics suggestive of outflowing gas (Murthy et al. in prep.), while there is no evidence of outflowing gas in the warm ionized phase (see the optical spectrum presented in \citealp{2006PhDT........60E}, and the public data from the CALIFA survey \footnote{http://califa.caha.es/}).

The time line presented in Fig.~\ref{timeline} summarizes all the information on these sources in the context of the scenario that we tested. The properties of the outflowing gas for PKS~B1934-63 and of the compact steep-spectrum radio galaxies described above support the scenario in which cold molecular gas might form within the outflow material through cooling of shock-heated gas. According to this scenario, PKS~B1718-649 and PKS~B1934-63 represent the earliest stages of the AGN-ISM interaction, when the outflowing gas is initially shock heated. This gas then starts to cool down in the post-shock region and can be detected as cold outflowing gas, as for 3C~305. Finally, the radio galaxy B2 0258+35 might be representative of the the final phases of the AGN-ISM interaction, when all the outflowing gas has completely cooled down. Considering the small amount of mass that is outflowing in the case of PKS~B1934-63, most of the mass may currently be in a hotter phase (e.g., gas at 10$^7$~K emitting in the X-ray band). 
  
We are aware that the radiative and kinematical age of a radio galaxy might differ, and that the radiative age indicates the age of the particles within the radio lobes rather than the real age of the source. However, the ages that we used are the only available estimates, and it has been shown that the radiative age of 3C~305 is representative of the age of the radio source \citep{1999A&A...345..769M}.  

The time line in Fig.~\ref{timeline} also includes the two compact steep-spectrum radio sources PKS~B1345+12 and IC~5063. These are two of the best examples of compact sources where extensive multi-wavelength studies found a multi-phase outflow (for PKS~B1345+12 see \citealp{2005A&A...444L...9M, 2011MNRAS.410.1527H, 2012A&A...541L...7D}; for IC~5063 see \citealp{2000AJ....119.2085O, Morganti2007, 2014Natur.511..440T, 2015A&A...580A...1M}). Unfortunately, no reliable estimate of the age of these radio sources is available. However, considering the aforementioned studies, they might represent an intermediate stage of an AGN-ISM interaction, and this justifies their location in the time line presented in Fig.~\ref{timeline}. 

\section{Conclusions}\label{discussion}

Compact and young radio galaxies are pivotal for our understanding of the feedback effect that radio sources have on their host galaxies. 
PKS~B1934-63 is a young ($\sim 1.6 \times10^3$ yr) radio source, and given it proximity (z=0.1824) and high radio power ($\rm{P}_{\rm 1.4GHz}=\rm{10}^{27.2}~\rm{W~Hz}^{-1}$), it is considered as an archetypal compact radio galaxy. 
 
The nuclear spectrum of PKS~B1934-63 shows double-peaked gas emission lines that have broad wings, indicating a complex kinematics for the warm ionized gas and outflowing gas.
The kinematical features of the intermediate and the very broad components, which are needed to model the broad part of the emission line profiles, and in particular their line width (FWHM of about 700 and 2000~\kms\ , respectively), clearly reflect non-gravitational motions of gas that is connected to an AGN-driven outflow.

We find that about $6.8\times10^6~\rm{M_{\odot}}$ of warm ionized gas is concentrated in the inner 500~pc of the host galaxy and only a small fraction of this gas is actually outflowing.
Most of the warm ionized gas is included in a structure that shows a smooth velocity gradient in the velocity-position diagram presented in Fig.~\ref{spectroastro}. Considering the rotating disk of warm ionized gas found by \cite{2016MNRAS.459.4259R} on larger scales, we tend to favor the hypothesis that this structure is a circum-nuclear disk with a radius of about 200 pc, which might constitute part of the gas reservoir from which the SMBH is fed (similar to what has been found, e.g., by~\citealp{Dumas2007}, \citealp{Hicks2013}, and \citealp{Maccagni2016}).

The results of our spectro-astrometry study show that when we assume a biconical geometry for the outflow, the spatial extent of the broad wings of the \OIII$\lambdaup$5007\AA\ line matches the diameter of the radio source (i.e., about 120 pc). This indicates that the outflow is likely driven by the expansion of the radio source jets, which is in line with what is commonly found in other compact and young radio sources \citep[see, e.g.,][]{2002AJ....123.2333O,2008MNRAS.387..639H,2014Natur.511..440T}.

Using the density-diagnostic diagram introduced by \cite{2011MNRAS.410.1527H}, we find a clear correlation between the FWHM of a component and its electron density (see Fig.~\ref{DDD}), which is in line with the findings of \cite{2011MNRAS.410.1527H} and \cite{Rose2017}. The warm ionized gas associated with the broad components reaches remarkably high electron densities ($10^{4.5}$-$10^{5.5}~\rm{cm^{-3}}$).
We attribute the broadening of the spectral lines to the interaction of the AGN with the ISM, and we argue that the FWHM-density relation that we find is mainly driven by the ability of the AGN-ISM interaction to compress the gas at different levels and increase its density. 

Estimating the gas densities with the transauroral line technique allowed us a reliable estimate of the properties of the warm outflowing ionized gas. We obtain low-mass outflow rates (i.e., highest values in the range 10$^{-3}$-10$^{-1}~\rm{M_{\odot}~yr^{-1}}$), and we find that only a small fraction of the available accretion power of the AGN is used to drive the outflow (i.e., maximum AGN feedback efficiency $F_{\rm kin}\sim$10$^{-3}\%$).
This does not match the results of \cite{2017A&A...601A.143F}, for example, who reported higher average AGN feedback efficiencies in their collection of ionized gas outflows (i.e., 0.16-0.3$\%$). However, the latter outflows lack a robust estimate of the gas density, which is always taken to be $\leq 10^{3}~$cm$^{-3}$, based on classical line ratio diagnostics or assumptions. 
It is worth mentioning that by adopting such densities, which are at least two order of magnitudes lower than the values we find, we would obtain an outflow mass rate and an AGN feedback efficiency that is compatible with the findings of \cite{2017A&A...601A.143F}. Instead, our results provide new evidence that AGN feedback occurs at a low efficiency, and  our results are  much rclose to the values reported by \cite{2016MNRAS.460..130V} for luminous type II AGN and values by \cite{Rose2017}, who used the same technique as we did here on a sample of local ULIRGs. 
This also highlights the importance of a fair comparison between the observed efficiencies and the cosmological models predictions \citep[see][for a discussion]{Harrison2018}.

Using the optical \SII$\lambdaup\lambdaup$6717,31/H$\alpha$, \OI$\lambdaup$6300/H$\alpha$, \OIII$\lambdaup\lambdaup$4958,5007/4363, and \HeII$\lambdaup$4686/H$\beta$ and the NIR \FeII$\rm{\lambdaup}$1.257$\upmu$m/Pa$\beta$ line-ratio diagnostics, we find that the AGN-ISM interaction drives shocks  that heat the gas within the ISM. In particular, the comparison between model grids and the observed optical line ratios in the BPT diagrams indicates that the broad components of the warm ionized gas are ionized by fast shocks with velocities  $v_{\rm s}\geq$400-500~\kms , possibly reaching a few thousand \kms . These velocities are compatible with the width of the broad components and suggest that shocks are a feasible mechanism to accelerate the warm ionized gas to high velocities.

By studying the \MgII\ absorption and the \Hdue\ emission lines, we find that the absorbing low-ionization ISM and warm molecular gas do not show signs of outflowing and/or kinematically disturbed gas.
In addition, past \HI\ observations of the nuclear region of the host galaxy did not find outflowing gas in the atomic phase \citep{2000A&A...362..426V}. 
This is at odds with other compact, steep-spectrum radio sources that are known to host ionized gas outflows, which usually show massive cold gas outflows \citep[][]{2005A&A...439..521M,2012A&A...541L...7D,2013Sci...341.1082M,2014Natur.511..440T,2015A&A...580A...1M}. 
By integrating our findings with information on other compact radio galaxies from the literature, we suggest that the lack of cold outflowing gas in PKS~B1934-63 might be explained by the fact that the shock-ionized outflowing gas did not have enough time to cool down given the young age of the radio source, and thus, the fact that the AGN-ISM interaction occurred recently. It is also possible that a fraction of the outflowing gas is in a hotter phase (e.g., T$\sim 10^7$K) that is missed by our observations. Our results strengthen the hypothesis that in general the AGN drives shocks within the ISM  during an AGN-ISM interaction, and cold gas is formed in situ within the outflowing material.
New observations probing the cold molecular gas (i.e., CO) in PKS~B1934-63 will be crucial for completing the picture on the presence and kinematics of the colder gas phase.

Our work shows that a systematic and detailed characterization of the multi-phase properties of the outflows in compact young radio galaxies offers the possibility to study the origin of cold gas within outflows. It also has the potential of shedding light on the relevance of AGN feedback operated by jets on galactic scales.

\begin{acknowledgements}
The authors thank the anonymous referee for the useful comments that improved the quality of the paper.
The research leading to these results has received funding from the European Research Council under the European Union's Seventh Framework Programme (FP/2007-2013) / ERC Advanced Grant RADIOLIFE-320745. Based on observations collected at the European Organisation for Astronomical Research in the Southern Hemisphere under ESO programme 87.B-0614A. 
\end{acknowledgements}

\bibliographystyle{aa}
\bibliography{biblio.bib}

\appendix 
\section{Stellar population and line fitting}\label{appendix1}

This appendix includes the results of the stellar population and emission line modeling. 
In the following we describe the fitting procedure for some of the spectral lines in more detail.

In the case of the \MgII$\lambdaup\lambdaup$2795,2802\AA\ doublet, the ISM \MgII\ absorption lines are superimposed on the emission lines by the AGN light. For each line of the doublet we used a Gaussian function to model the AGN emission and a Gaussian function to model the ISM absorption (see Fig.~\ref{MgIIfit}). The separation between the two emission and absorption components was fixed according to the laboratory rest wavelengths of the doublet. In addition, we fixed their sigma to be the same. We find that the \MgII\ emission is fit by a Gaussian with an FWHM=1245$\pm$99\kms\ and a velocity shift v=$-53\pm 35$\kms\ , while the absorption component has an FWHM=200$\pm 38$\kms\ and a velocity shift v=14$\pm 36$\kms.

In the case of the \NII\ and \OI\ lines, we fixes the \NII$\lambdaup$6584/$\lambdaup$6548 and the \OI$\lambdaup$6300/$\lambdaup$6363 line ratios to be equal to 3 according to atomic physics \citep{2006agna.book.....O}.  The \OII$\lambdaup$7319/$\lambdaup$7330 line ratio was fixed to be 1.24 because it was found not to vary with density. 
The \OII$\lambdaup$3729/$\lambdaup$3726 and the \SII$\lambdaup$6717/$\lambdaup$6731 ratios were limited within their theoretical values at low and high densities \citep{2006agna.book.....O}. We also limited the \SII$\lambdaup$4069/$\lambdaup$4076 ratio to be in the range 3.01-3.28 according to the calculation made by \cite{Rose2017} at different densities.

The \OI$\lambdaup$6300\AA\ line shows a different profile compared to all the other forbidden emission lines, which might be due to contamination from the \SIII$\lambdaup$6312\AA\ line. Considering the flux of the \SIII$\lambdaup$9531\AA\ emission line, we can predict the minimum total flux and peak emission expected for the \SIII$\lambdaup$6312\AA\ line. To do this, we assumed the lower value that the \SIII\ 9531/6312 line ratio reaches in the high temperature regime \citep[see][]{2006agna.book.....O}. 
We find that we expect to detect the emission of the \SIII$\lambdaup$6312\AA\ line in the nuclear spectrum of PKS~B1934-63. We predict that the \SIII$\lambdaup$6312\AA\ line has a total flux higher than 17.5$\times10^{-17} \rm{erg~s^{-1}cm^{-2}}$ and a line peak (assuming the four kinematical components of the \OIII\ model) greater than about 3$\times10^{-17} \rm{erg~s^{-1}cm^{-2}\AA^{-1}}$.
For this reason, when we fit the \OI$\lambdaup\lambdaup$6300-63\AA\ doublet, we also included an \SIII$\lambdaup$6312\AA\ component that is well reproduced using only the intermediate component of the \OIII\ model (see Fig.~\ref{OI6300}). 

In the fit of the \OII$\lambdaup\lambdaup$7319,30\AA\ we also introduced an additional Gaussian component to take into account the emission of the \OII$\lambdaup$7381\AA\ emission in the red wing of the doublet. In the same way, we take into account the emission of the \Hdue\ S(4)1-0 line in the fit of the Pa$\alpha$ line profile.

\begin{figure*}[]
\centering
\includegraphics[width=\hsize, keepaspectratio]{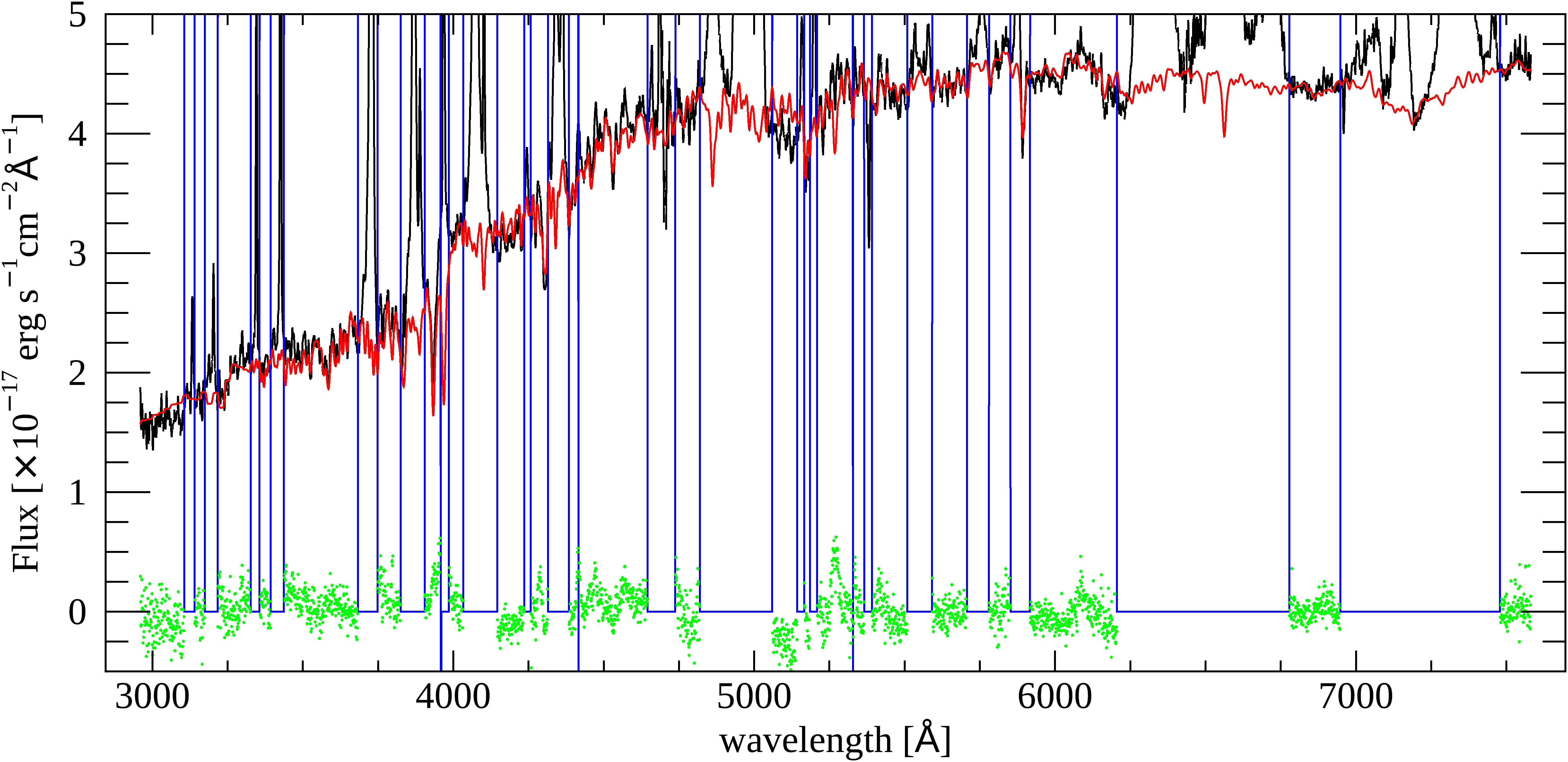}
\caption{ Nuclear spectrum (black solid line) and best-fit model (red solid line) for the continuum emission and residual (green points). The masked regions that correspond to the emission lines are indicated in blue. The observed spectrum is smoothed with a box of 3\AA . } 
\label{SPfitting}
\end{figure*}

\begin{table*}
\centering 
\begin{tabular} { p{2cm} p{1.5cm} p{1.5cm} p{1.5cm} p{1.5cm} p{1.5cm} } 
\hline              
\noalign{\smallskip}
        Line ID &   $\rm{\lambdaup}$ ($\AA\ $) & Flux$_{\rm 1N}$  & Flux$_{\rm 2N}$ & Flux$_{\rm I}$  & Flux$_{\rm VB}$ \\     
\hline\hline
\noalign{\smallskip} 
 \OII 3726 &  3726.03  & 34.7 & 46.8 & 91.1 & 28.7  \\  
\noalign{\smallskip}
\OII 3729 &  3728.82 & 36.9 & 48.1 & 39.9  & 7.10    \\ 
\noalign{\smallskip}
\SII 4068 & 4068.60 & 1.40 & 2.50  & 52.0 & 34.0    \\ 
\noalign{\smallskip}
\SII 4076 & 4076.35 & 0.42   & 0.84  & 17.0 & 11.0   \\
\noalign{\smallskip}
H$\gamma$ & 4340.50 & 6.95   & 6.20   & 27.6  & 18.1      \\
\noalign{\smallskip}
                     &               & 7.23  & 6.41   & 24.3  & 9.51$\pm$2.56      \\
\noalign{\smallskip}
\OIII 4636 & 4363.20 &  5.80  & 3.94   & 10.5  & 22.3     \\
\noalign{\smallskip}
                 &               &  5.48   & 3.68     & 14.7    & --       \\
\noalign{\smallskip}
\HeII 4686 & 4685.7 & 5.41  & 4.90  & 4.19$\pm$0.74 & -- \\
\noalign{\smallskip}
H$\beta$ & 4861.33 & 21.9 & 24.1 & 59.0 & 16.6$\pm$7.6   \\
\noalign{\smallskip}
\OIII 5007 & 5006.84 & 297  & 297 & 218 & 197  \\
\noalign{\smallskip}
\OI 6300 & 6300.30 &  54.4 & 65.2 & 240 & 211   \\
\noalign{\smallskip}
\SIII 6312 & 6312.10 & --  & -- & 34.0  & --  \\
\noalign{\smallskip}
H$\alpha$ & 6562.80 & 99.4$\pm$10.3 & 106$\pm$11 & 306$\pm$32 & 58.0$\pm$10.1     \\
\noalign{\smallskip}
\NII 6584 & 6583.41 & 95.7$\pm$10.1  & 106$\pm$11 & 188$\pm$20 & 295$\pm$32     \\
\noalign{\smallskip}
\SII 6717 & 6716.47 & 40.7  & 41.5 & 41.5 & 20.0$\pm$13.2     \\
\noalign{\smallskip}
\SII 6731 & 6730.85 & 36.7  & 39.9 & 92.1 & 14.7$\pm$5.46     \\
\noalign{\smallskip}
\OII 7318,19 & 7320.10 & 4.07  & 5.03  & 51.8 & 92.7    \\
\noalign{\smallskip}
\OII 7330,31 & 7330.20 & 3.28  & 4.05  &  41.7 &74.7   \\
\noalign{\smallskip}
\SIII 9531 & 9530.6 & 59.0  & 61.1  & 83.0 & 56.1$\pm$7.0  \\
\noalign{\smallskip}
Pa$\alpha$ & 18751.01 & 17.7  & 18.0  & 29.0  & 66.8    \\
\hline
\end{tabular}
\caption{Ffluxes of the four kinematical components of each fitted line. The fluxes are given in units of $10^{-17}~$erg~s$^{-1}$cm$^{-2}$. The first and second column indicate the emission line ID and rest frame wavelength, respectively. For the H$\gamma$ and the \OIII$\lambdaup$4363\AA\ lines we report two sets of fluxes. These lines are blended, and the first row reports the fluxes obtained by modeling both lines with the \OIII\ model, while the second row reports the fluxes obtained by modeling the H$\gamma$ line with the H$\beta$ model and the \OIII$\lambdaup$4363\AA\ line with the \OIII\ model. When not indicated, the error on the reported flux is 10$\%$.}
\label{Table_linefluxes}
\end{table*}

\begin{figure}[]
\centering
\includegraphics[width=\hsize, keepaspectratio]{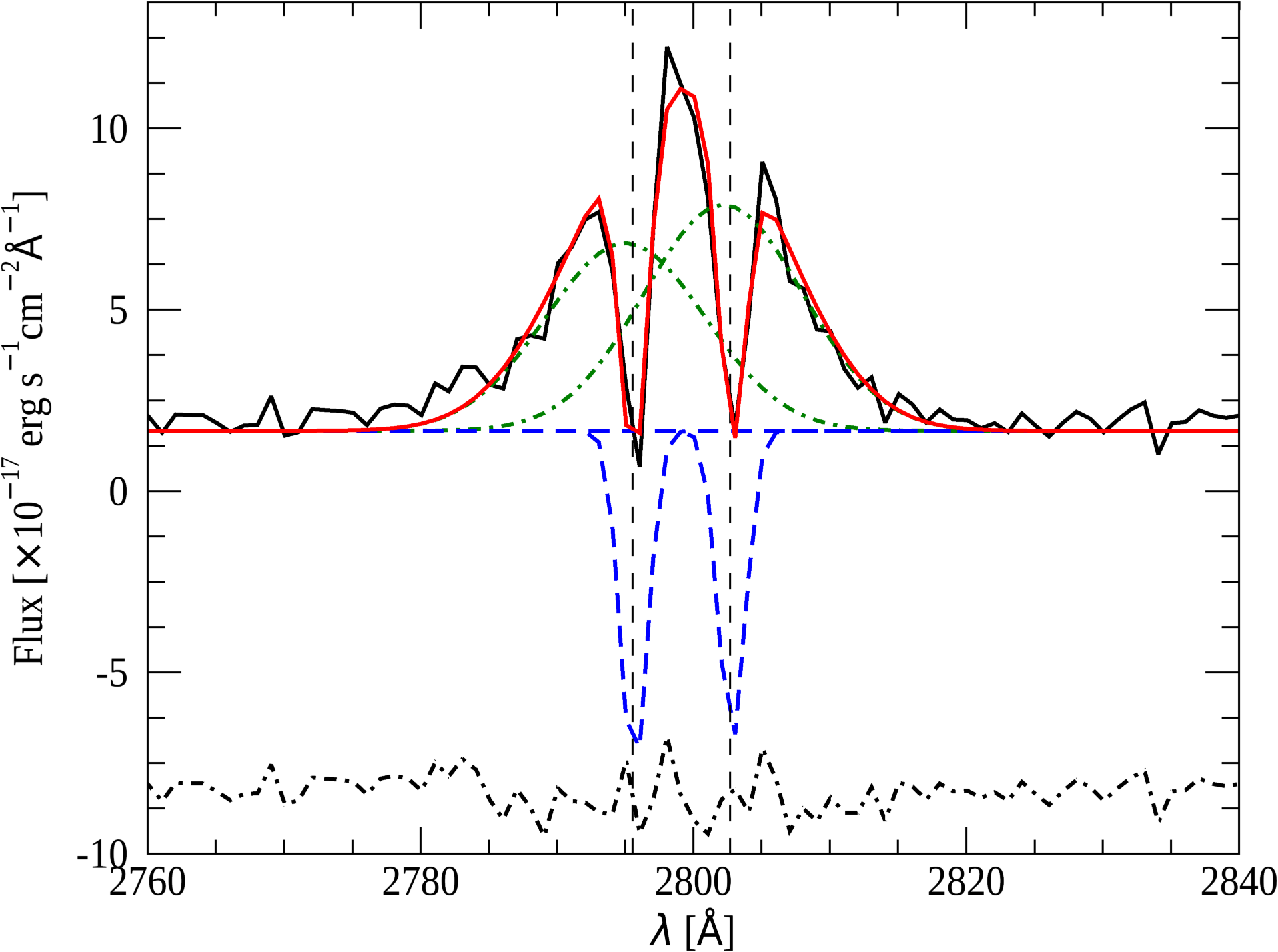}
\caption{{\rm Mg}\,{\rm \scriptsize II}$\lambdaup\lambdaup$2795,2802\AA\ lines (black solid line) and best-fit model (red solid line). Each line is fit with one Gaussian component to take into account the emission (green dot-dashed line) and one Gaussian component to take into account the absorption (blue dashed line). The residuals of the fit are normalized and plotted below the spectrum (black dot-dashed line). The vertical dashed lines mark the restframe wavelength of the fitted emission lines.} 
\label{MgIIfit}
\end{figure}

\begin{figure}[]
\centering
\includegraphics[width=\hsize, keepaspectratio]{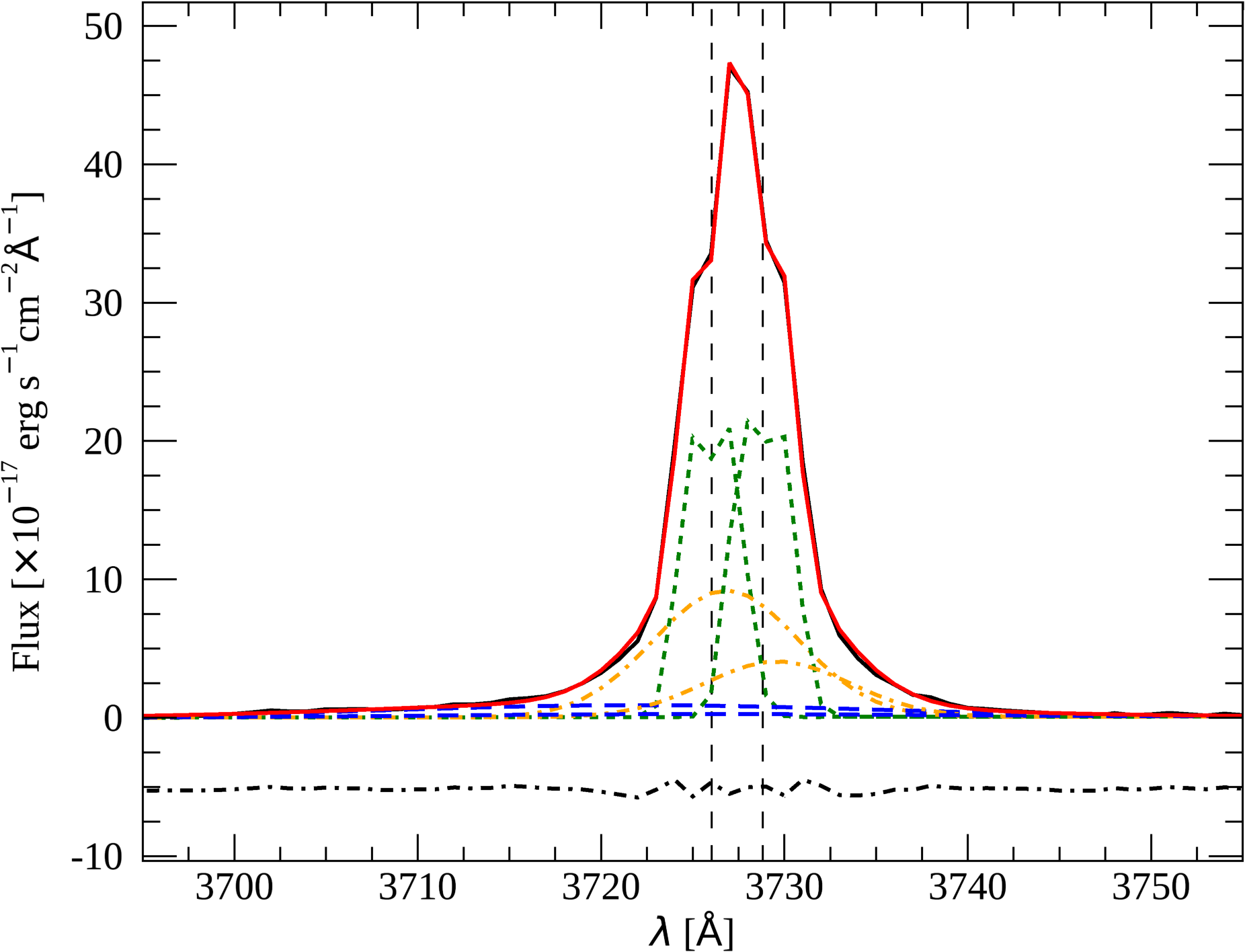}
\caption{{\rm[O}\,{\rm \scriptsize II}{\rm]}$\rm{\lambdaup\lambdaup}$3726,29\AA\ emission lines (black solid line) and best-fit model (red solid line). Each line is modeled using the {\rm[O}\,{\rm \scriptsize III}{\rm]} model, the 1N and 2N components are plotted together with the green dotted line, the I component is shown with the golden dot-dashed line, and the VB component with the blue dashed line.
The residuals of the fit are normalized and plotted below the spectrum (black dot-dashed line). The vertical dashed lines mark the restframe wavelength of the fitted emission lines.} 
\label{OII3727,29}
\end{figure}

\begin{figure}[]
\centering
\includegraphics[width=\hsize, keepaspectratio]{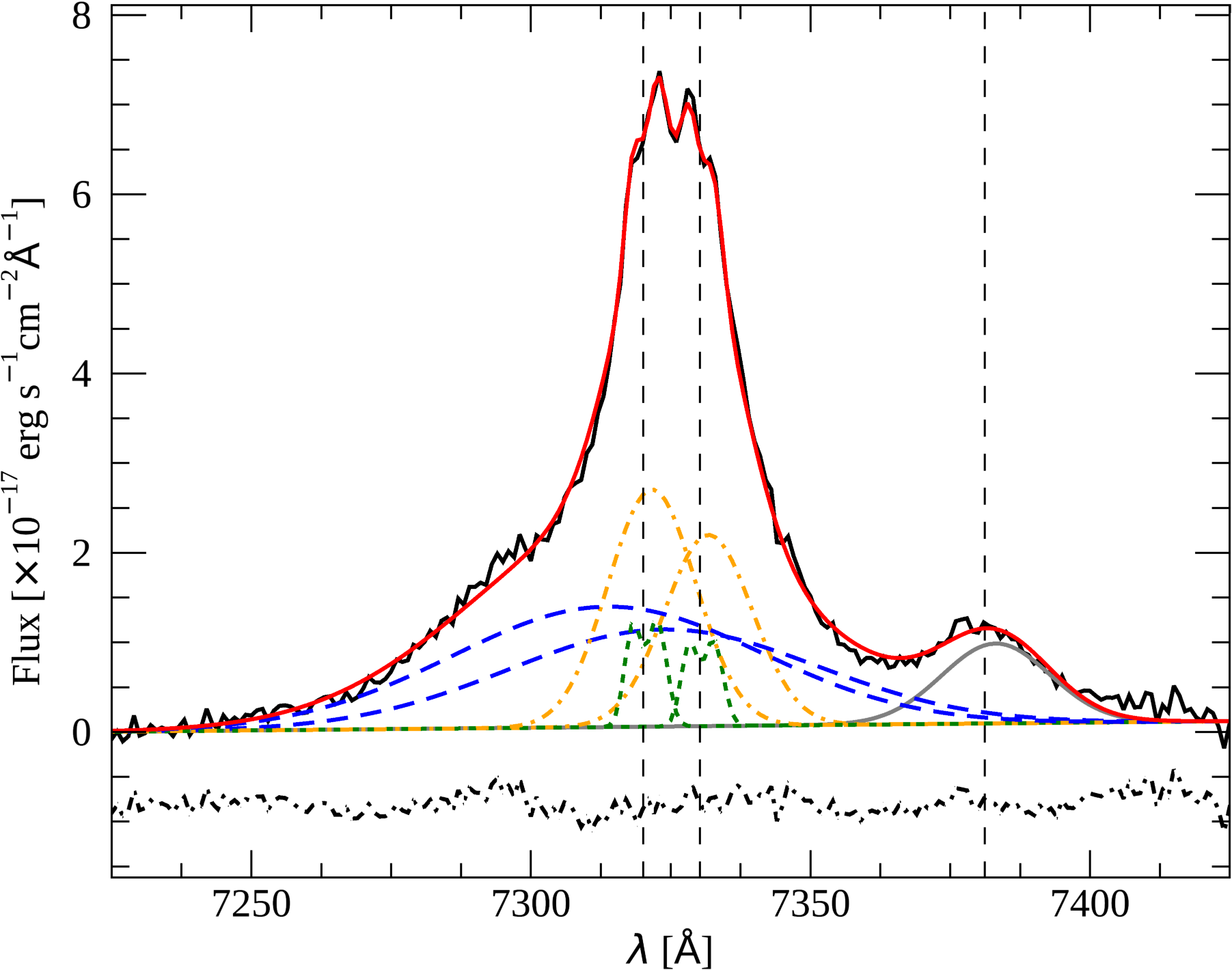}
\caption{{\rm[O}\,{\rm \scriptsize II}{\rm]}$\rm{\lambdaup\lambdaup}$7319,30\AA\ and  {\rm[O}\,{\rm \scriptsize II}{\rm]}$\lambdaup$7381\AA\  emission lines (black solid line) and best-fit model (red solid line). Each line of the {\rm[O}\,{\rm \scriptsize II}{\rm]}$\rm{\lambdaup\lambdaup}$7320-30\AA\ doublet is modeled using the {\rm[O}\,{\rm \scriptsize III}{\rm]} model, the 1N and 2N components are plotted together with the green dotted line, the I component is shown with the golded dot-dashed line, and the VB component with the blue dashed line. The {\rm[O}\,{\rm \scriptsize II}{\rm]}$\lambdaup$7381\AA\ line is modeled with a single Gaussian function indicated with the gray solid line.
The residuals of the fit are normalized and plotted below the spectrum (black dot-dashed line). The vertical dashed lines mark the restframe wavelength of the fitted emission lines.} 
\label{OII7320-30}
\end{figure}

\begin{figure}[]
\centering
\includegraphics[width=\hsize, keepaspectratio]{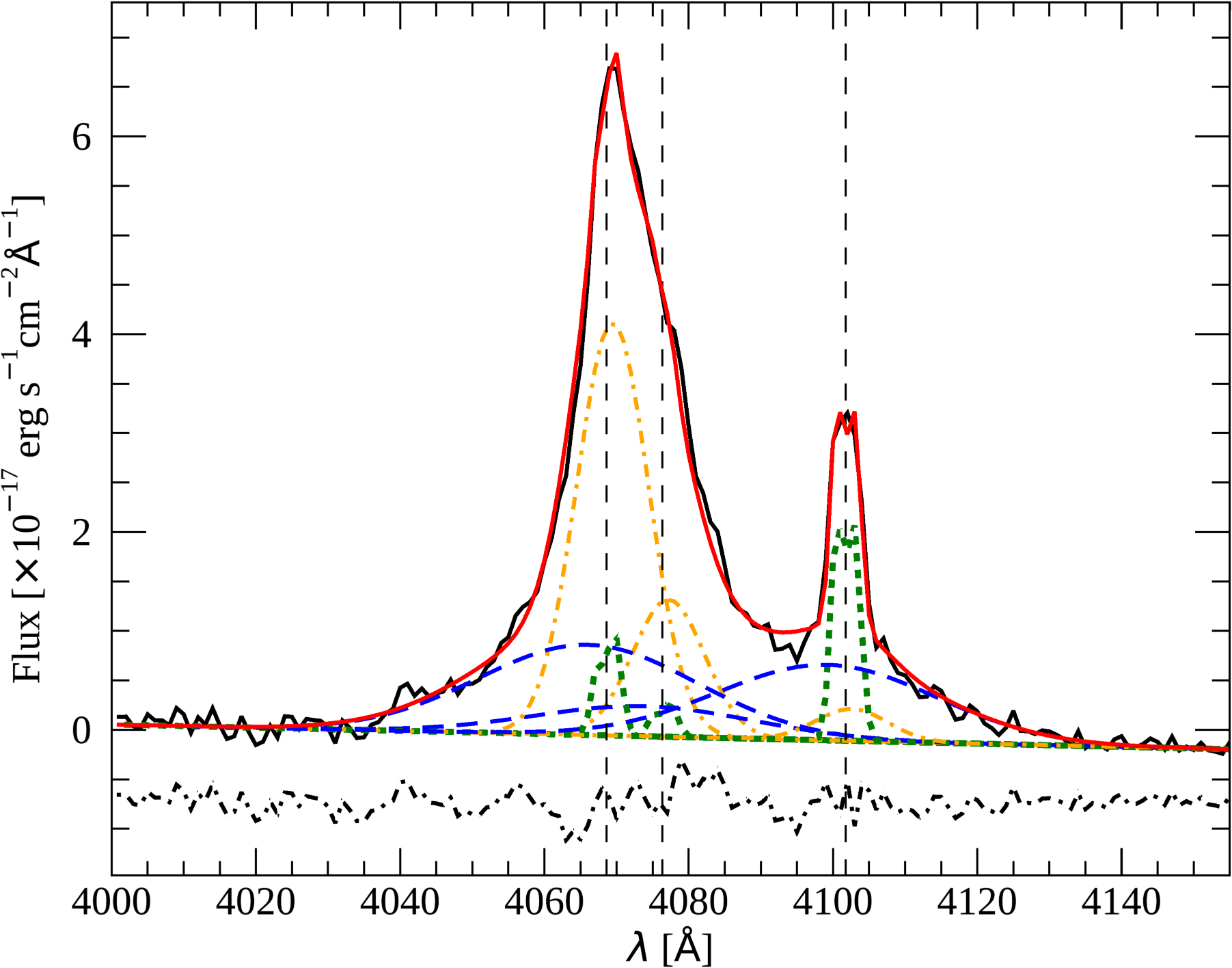}
\caption{{\rm[S}\,{\rm \scriptsize II}{\rm]}$\rm{\lambdaup\lambdaup}$4069,76\AA\ and H$\delta$ emission lines (black solid line) and best-fit model (red solid line). Each line is modeled using the {\rm[O}\,{\rm \scriptsize III}{\rm]} model, the 1N and 2N components are plotted together with the green dotted line, the I component is shown with the golden dot-dashed line, and the VB component with the blue dashed line.
The residuals of the fit are normalized and plotted below the spectrum (black dot-dashed line). The vertical dashed lines mark the restframe wavelength of the fitted emission lines.
} 
\label{SII4068,76}
\end{figure}

\begin{figure}[]
\centering
\includegraphics[width=\hsize, keepaspectratio]{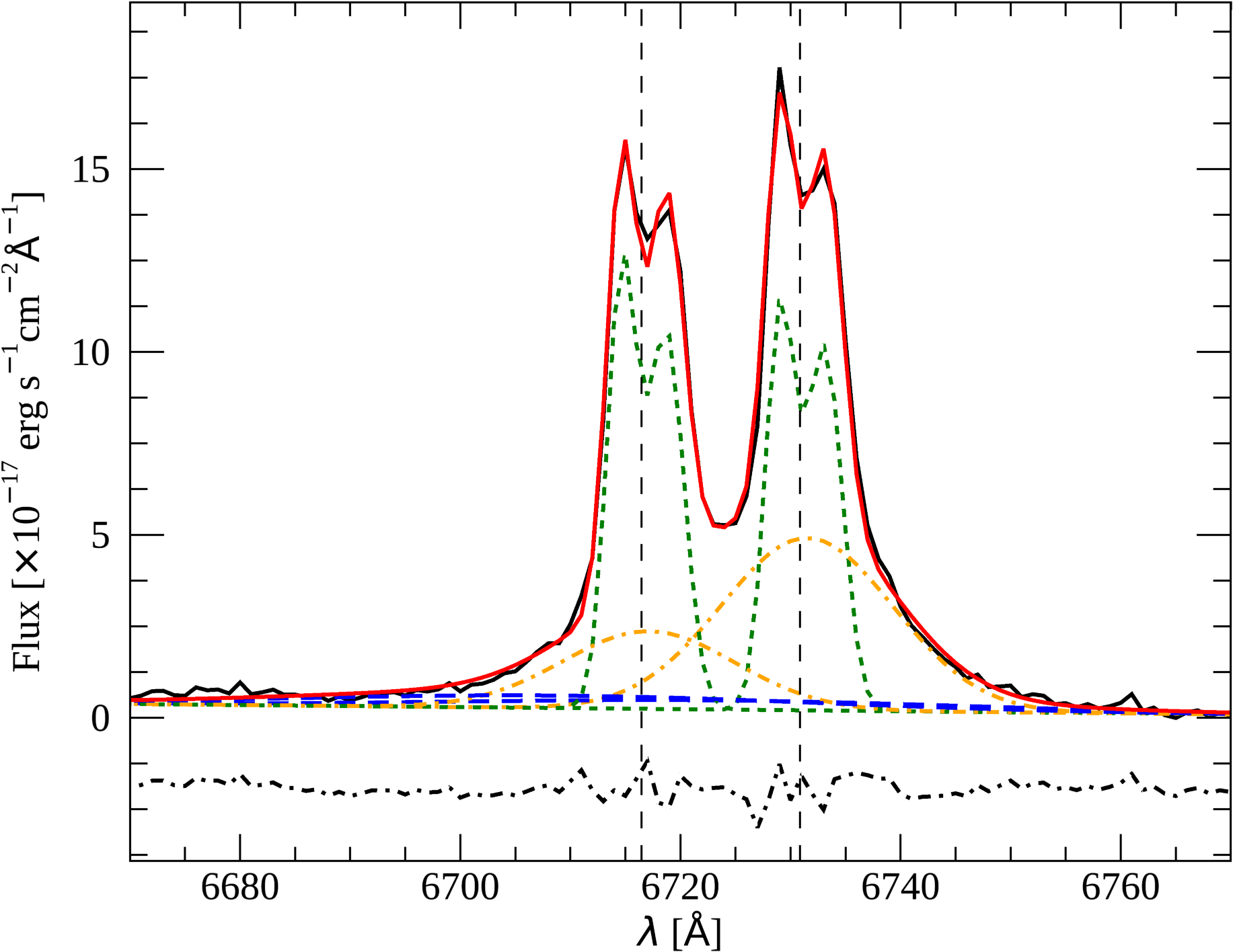}
\caption{{\rm[S}\,{\rm \scriptsize II}{\rm]}$\rm{\lambdaup\lambdaup}$6717,31\AA\ emission lines (black solid line) and best-fit model (red solid line). Each line is modeled using the {\rm[O}\,{\rm \scriptsize III}{\rm]} model, the 1N and 2N components are plotted together with the green dotted line, the I component is shown with the golden dot-dashed line, and the VB component with the blue dashed line.
The residuals of the fit are normalized and plotted below the spectrum (black dot-dashed line). The vertical dashed lines mark the restframe wavelength of the fitted emission lines.} 
\label{SII6716,31}
\end{figure}

\begin{figure}[]
\centering
\includegraphics[width=\hsize, keepaspectratio]{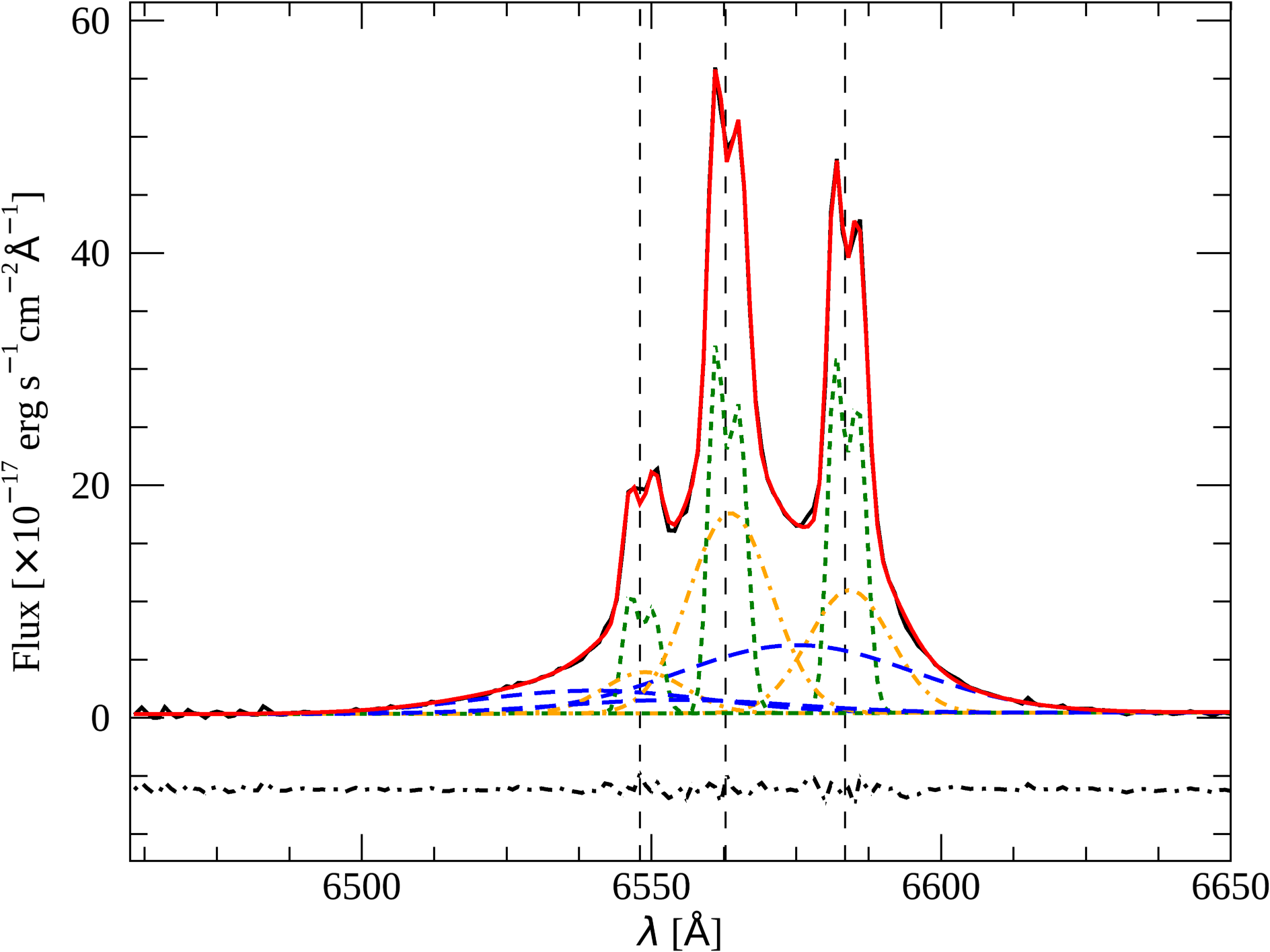}
\caption{ H$\alpha$ and {\rm[N}\,{\rm \scriptsize II}{\rm]}$\rm{\lambdaup\lambdaup}$6548-84\AA\ emission lines (black solid line) and best-fit model (red solid line). Each line is modeled using the {\rm[O}\,{\rm \scriptsize III}{\rm]} model, the 1N and 2N components are plotted together with the green dotted line, the I component is shown with the golded dot-dashed line, and the VB component with the blue dashed line.
The residuals of the fit are normalized and plotted below the spectrum (black dot-dashed line). The vertical dashed lines mark the restframe wavelength of the fitted emission lines.} 
\label{NII}
\end{figure}

\begin{figure}[]
\centering
\includegraphics[width=\hsize, keepaspectratio]{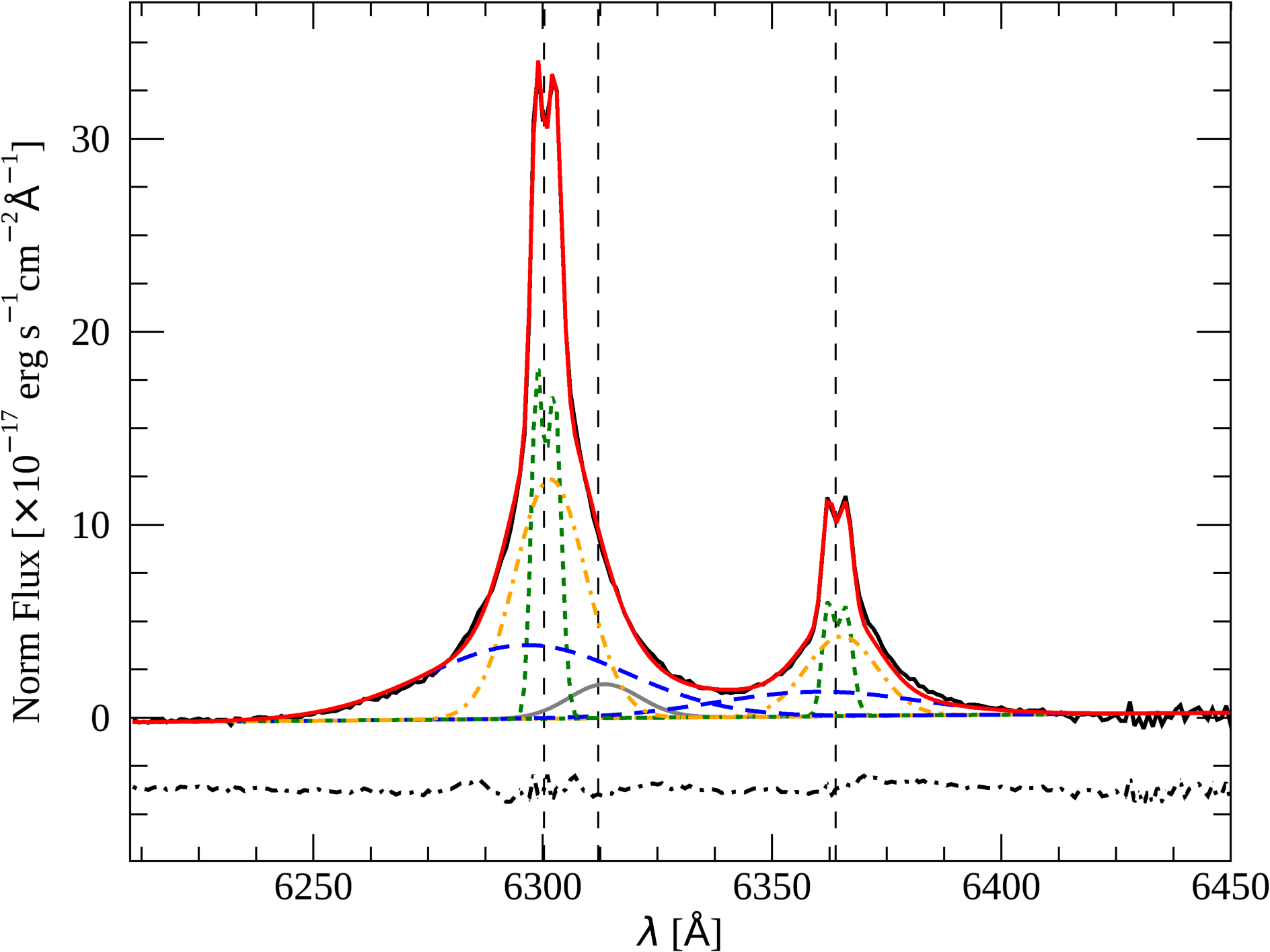}
\caption{{\rm[O}\,{\rm \scriptsize I}{\rm]}$\rm{\lambdaup\lambdaup}$6300-63\AA\ and {\rm[S}\,{\rm \scriptsize III}{\rm]}$\lambdaup$6312\AA\ emission lines (black solid line) and best-fit model (red solid line). Each line of the {\rm[O}\,{\rm \scriptsize I}{\rm]}$\rm{\lambdaup\lambdaup}$6300-63\AA\ doublet is modeled using the {\rm[O}\,{\rm \scriptsize III}{\rm]} model, the 1N and 2N components are plotted together with the green dotted line, the I component is shown with the golded dot-dashed line, and the VB component with the blue dashed line. The {\rm[S}\,{\rm \scriptsize III}{\rm]}$\lambdaup$6312\AA\ line is modeled with a single Gaussian function indicated with the gray solid line. The residuals of the fit are normalized and plotted below the spectrum (black dot-dashed line). The vertical dashed lines mark the restframe wavelength of the fitted emission lines.  } 
\label{OI6300}
\end{figure}

\begin{figure}[]
\centering
\includegraphics[width=\hsize, keepaspectratio]{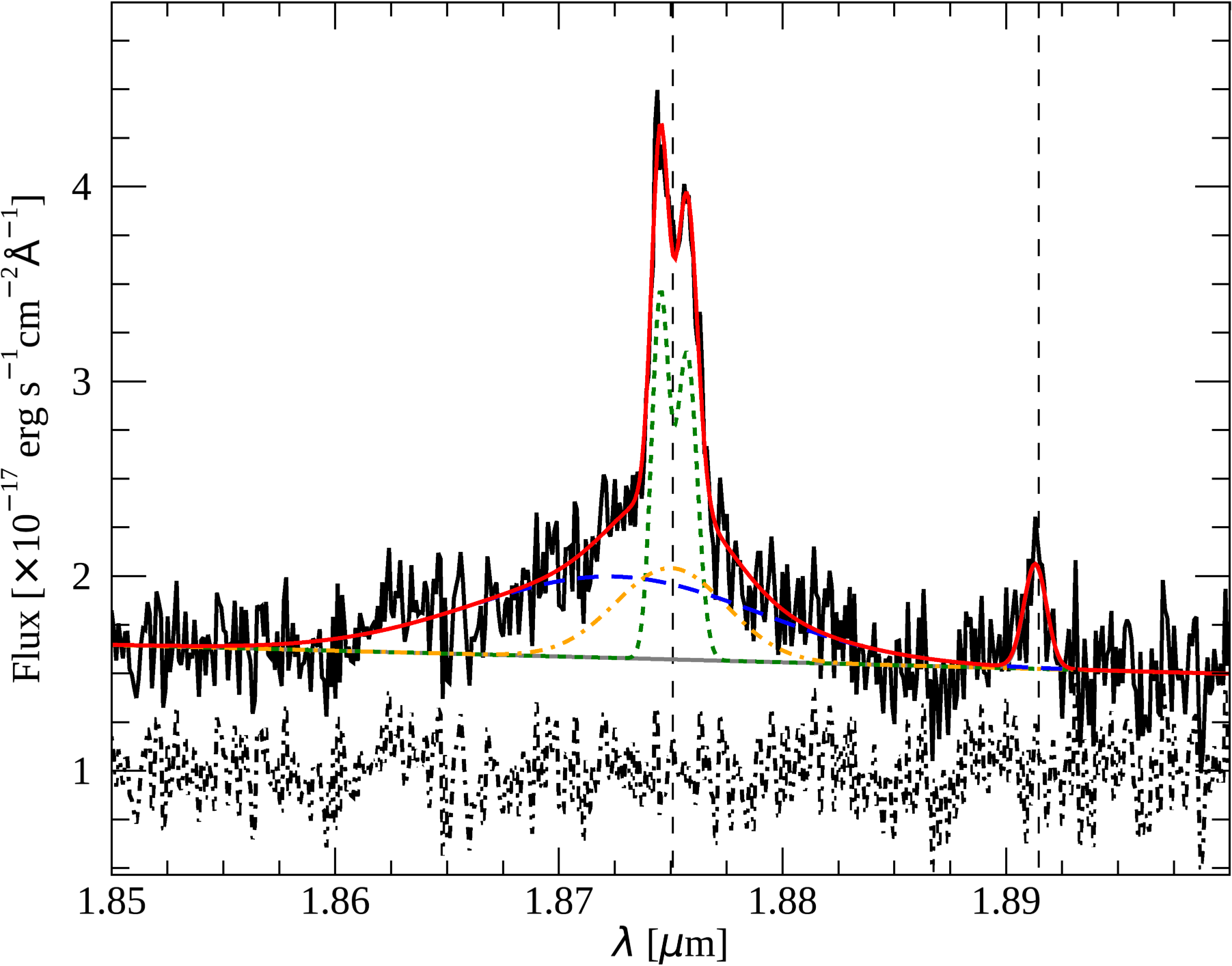}
\caption{Pa$\alpha$ and \Hdue\ S(4)1-0  emission lines (black solid line) and best-fit model (red solid line). The Pa$\alpha$ line is modeled using the {\rm[O}\,{\rm \scriptsize III}{\rm]} model, the 1N and 2N components are plotted together with the green dotted line, the B component is shown with the golden dot-dashed line, and the VB component with the blue dashed line. The \Hdue\ S(4)1-0  is modeled with a single Gaussian function indicated with the gray solid line. The residuals of the fit are normalized and plotted below the spectrum (black dot-dashed line). The vertical dashed lines mark the restframe wavelength of the fitted emission lines.} 
\label{Palpha}
\end{figure}

\begin{figure}[]
\centering
\includegraphics[width=\hsize, keepaspectratio]{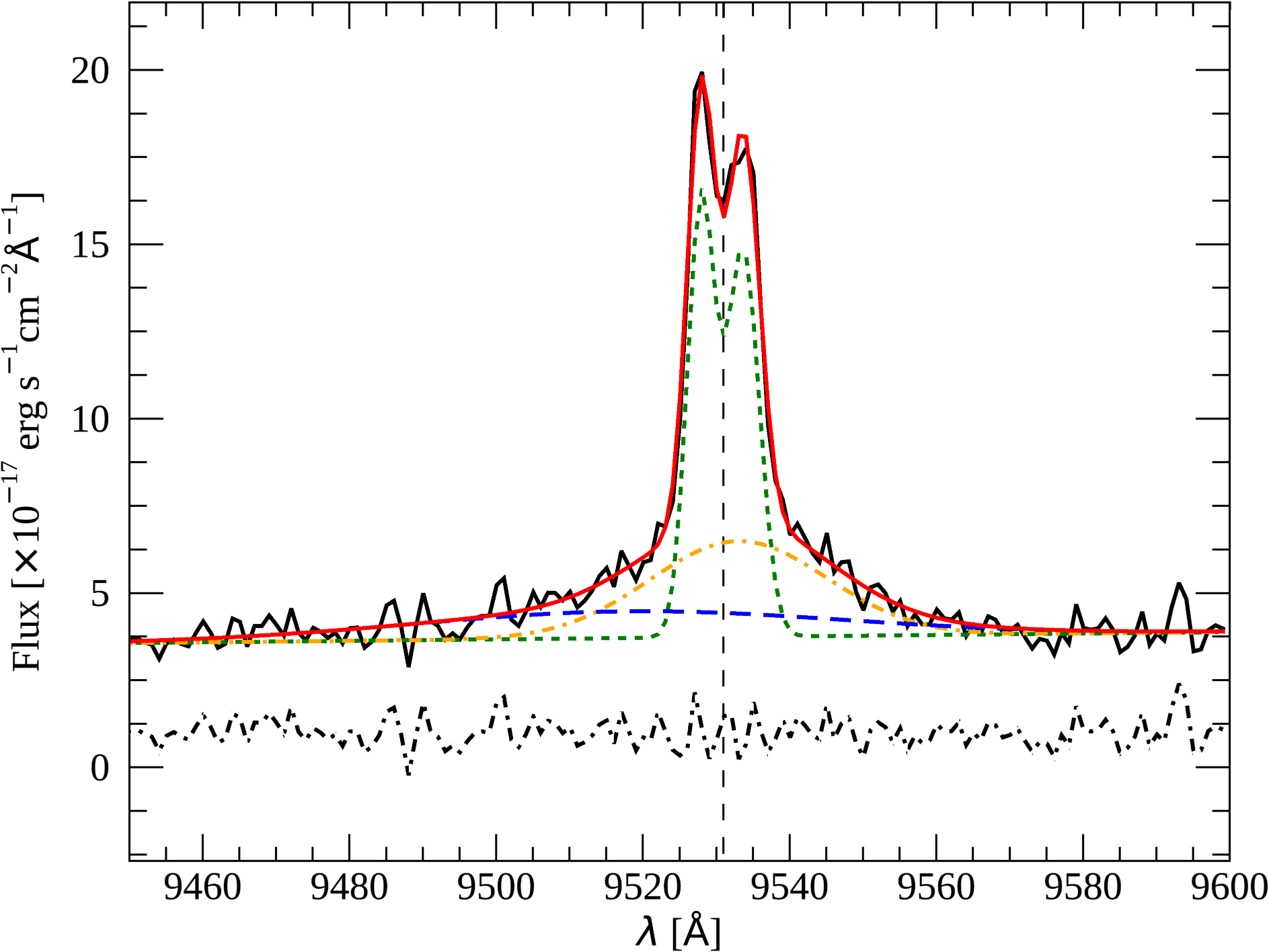}
\caption{{\rm[S}\,{\rm \scriptsize III}{\rm]}$\lambdaup$9531\AA\ emission line (black solid line) and best-fit model (red solid line). The {\rm[S}\,{\rm \scriptsize III}{\rm]}$\lambdaup$9531\AA\ line is modeled using the {\rm[O}\,{\rm \scriptsize III}{\rm]} model, the 1N and 2N components are plotted together with the green dotted line, the I component is shown with the golden dot-dashed line, and the VB component with the blue dashed line. The residuals of the fit are normalized and plotted below the spectrum (black dot-dashed line). The vertical dashed lines mark the restframe wavelength of the fitted emission line.} 
\label{Palpha}
\end{figure}

\begin{figure}[]
\centering
\minipage{0.5\textwidth}
\includegraphics[width=\hsize, keepaspectratio]{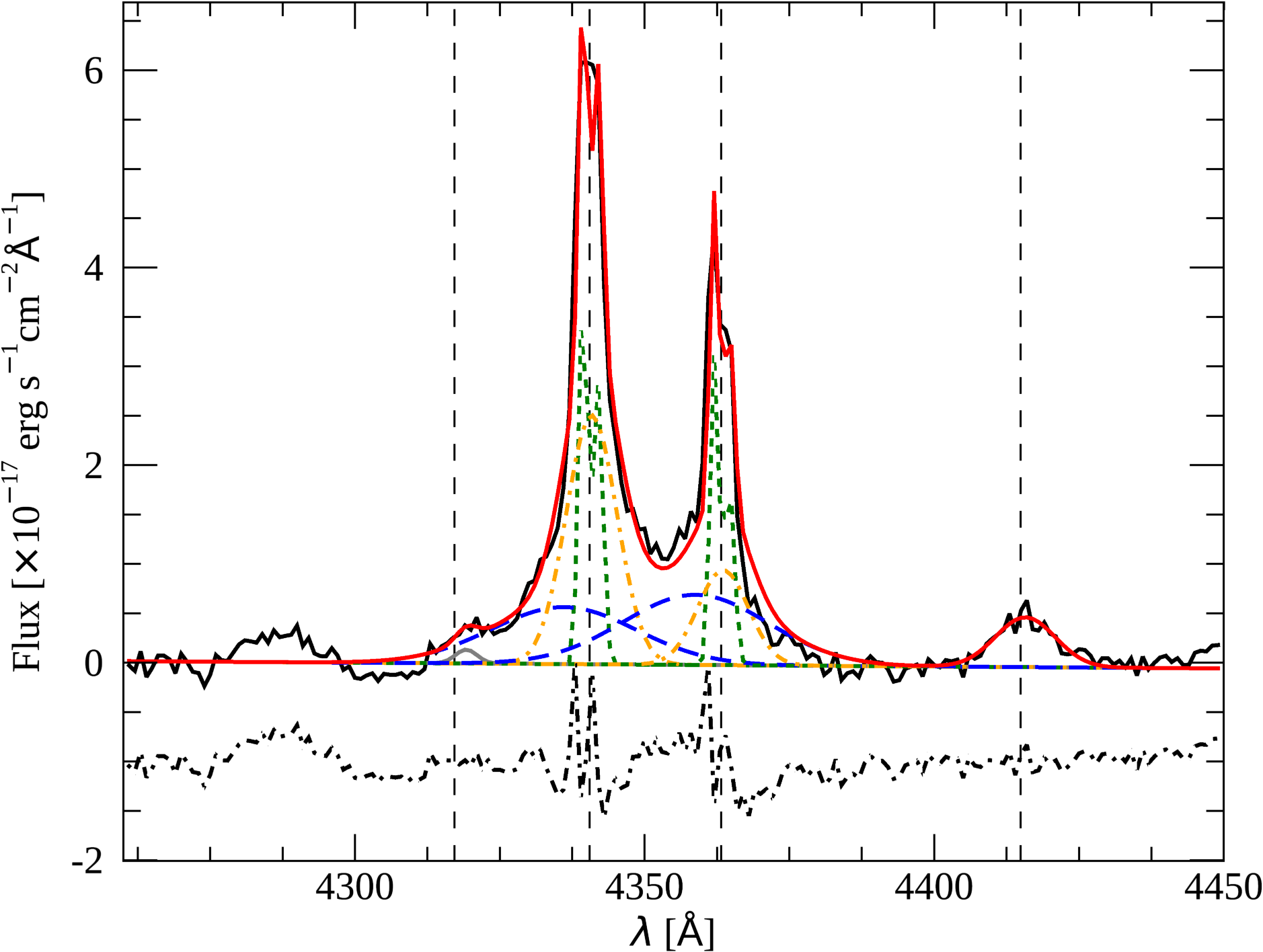}
\endminipage\hfill
\minipage{0.5\textwidth}
\includegraphics[width=\hsize, keepaspectratio]{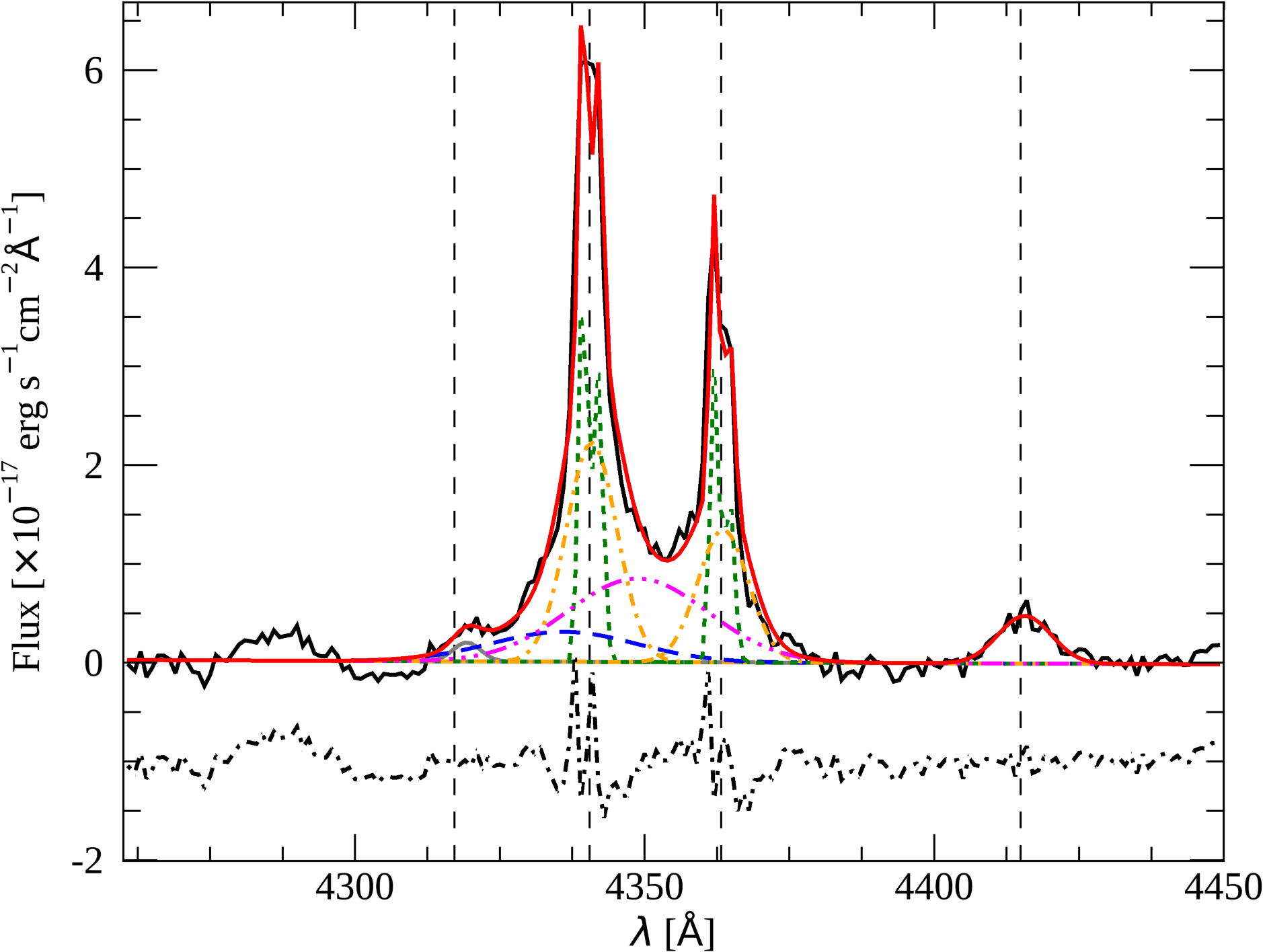}
\endminipage\hfill

\caption{H$\gamma$+{\rm[O}\,{\rm \scriptsize III}{\rm]}$\lambdaup$4363\AA\ emission lines (black solid line) and best-fit model (red solid line). The fitting procedure includes the modeling of two additional emission lines (gray solid line), most likely the \OII$\lambdaup$4317.2\AA\ and the \OII$\lambdaup$4414.9\AA\ lines, using a single Gaussian function. The residuals of the fit are normalized and plotted below the spectrum (black dot-dashed line). The vertical dashed lines mark the restframe wavelength of the fitted emission lines.
The \OIII$\lambdaup$4363\AA\ line is modeled using the \OIII\ model, and the H$\gamma$ line is modeled using both the \OIII\ model (\textit{top panel}) and the H$\beta$ model (\textit{bottom panel}). The 1N and 2N components are plotted together with the green dotted line, the I component is shown with the golden dot-dashed line, and the VB component with the blue dashed line. The magenta triple dot-dashed line indicates the broad redshifted component of the H$\beta$ model.} 
\label{Hg_OIII}
\end{figure}

\begin{figure}[]
\centering
\includegraphics[width=\hsize, keepaspectratio]{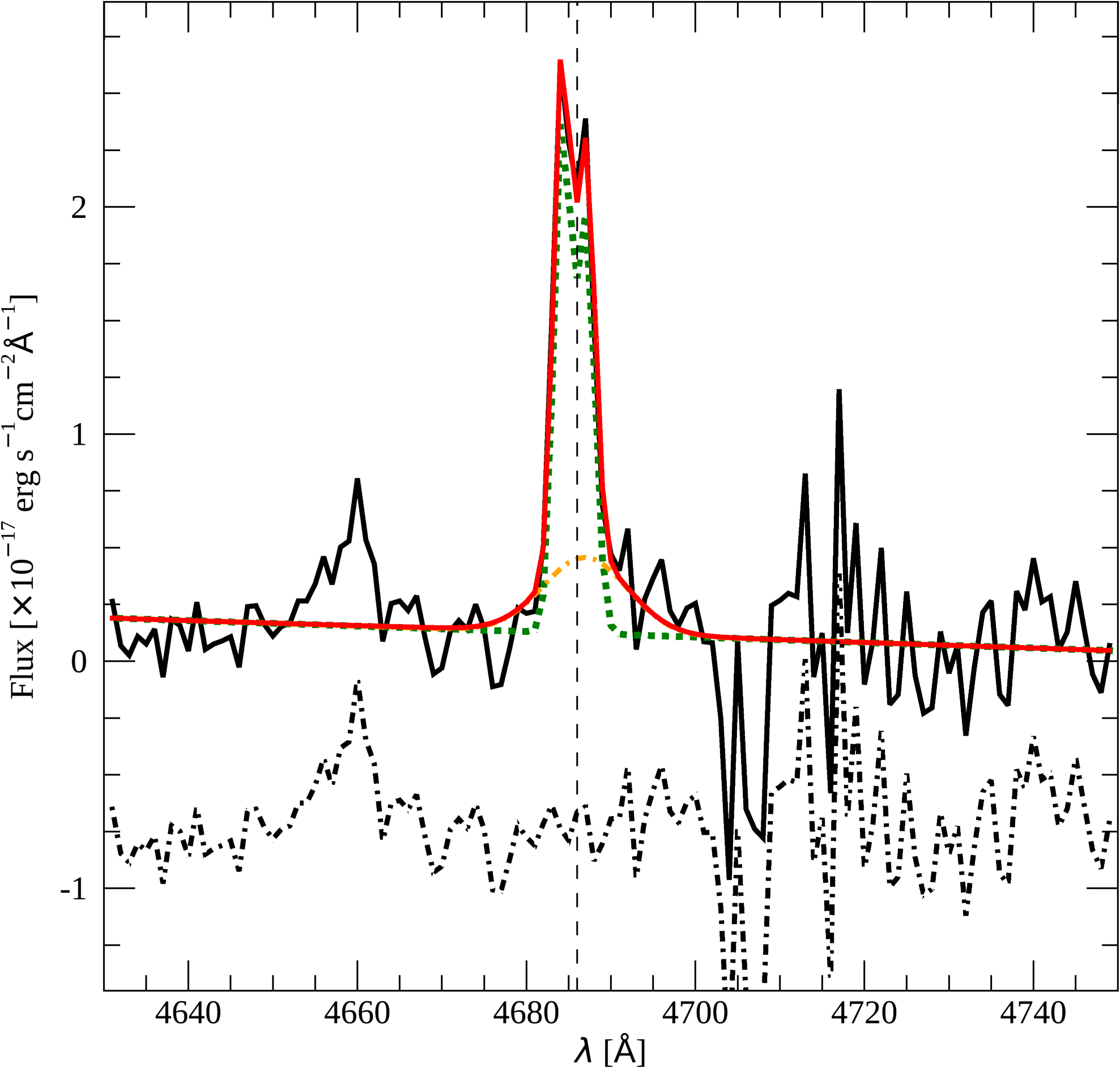}
\caption{\HeII$\lambdaup$4686\AA\ emission line (black solid line) and best-fit model (red solid line). The line is modeled using the {\rm[O}\,{\rm \scriptsize III}{\rm]} model, the 1N and 2N components are plotted together with the green dotted line, the I component is shown with the golden dot-dashed line. There is no evidence of a VB component. The residuals of the fit are normalized and plotted below the spectrum (black dot-dashed line). The vertical dashed lines mark the restframe wavelength of the fitted emission line.} 
\label{HeII}
\end{figure}

\begin{figure}[]
\centering
\minipage{0.455\textwidth}
\includegraphics[width=\hsize, keepaspectratio]{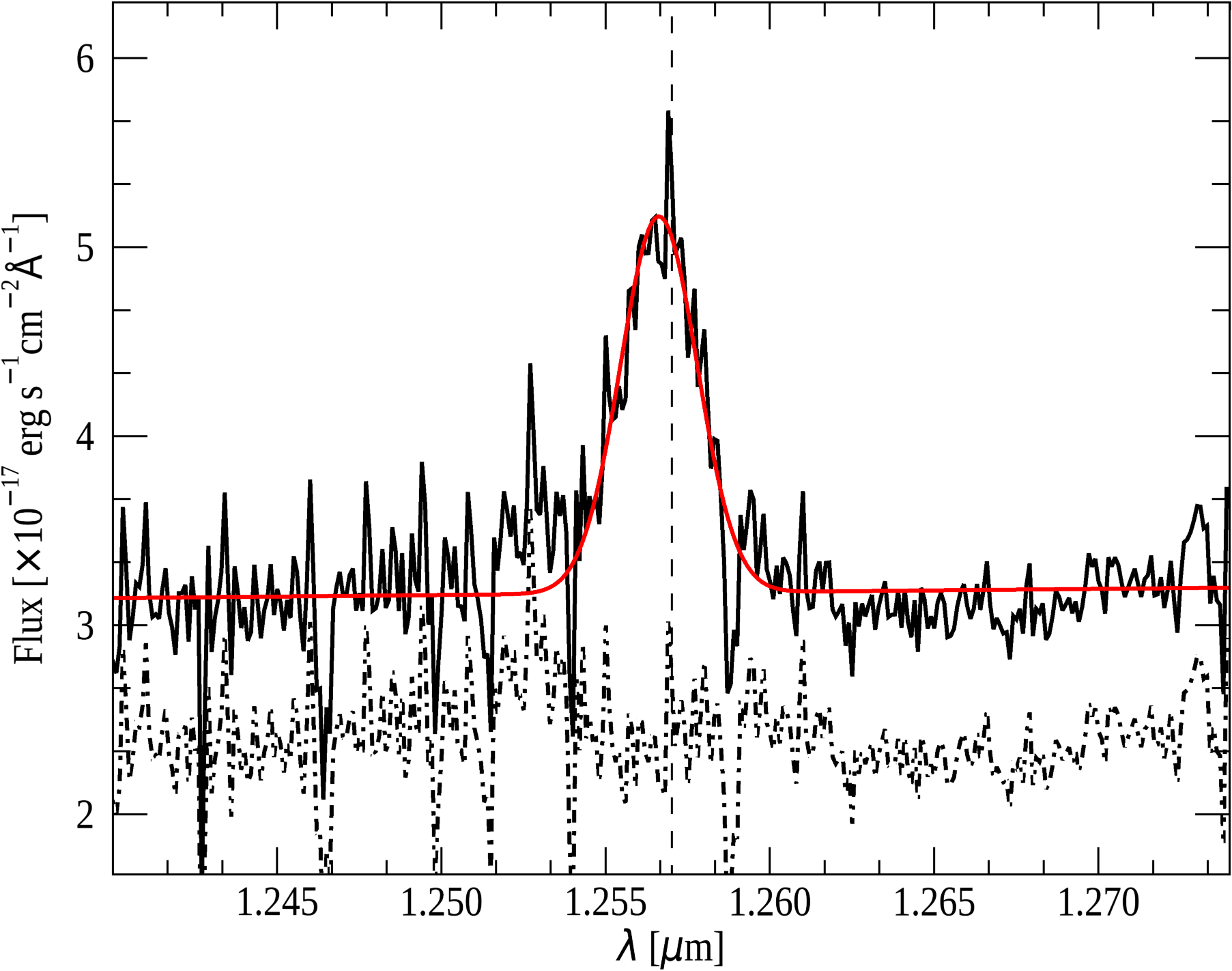}
\endminipage\hfill
\minipage{0.5\textwidth}
\includegraphics[width=\hsize, keepaspectratio]{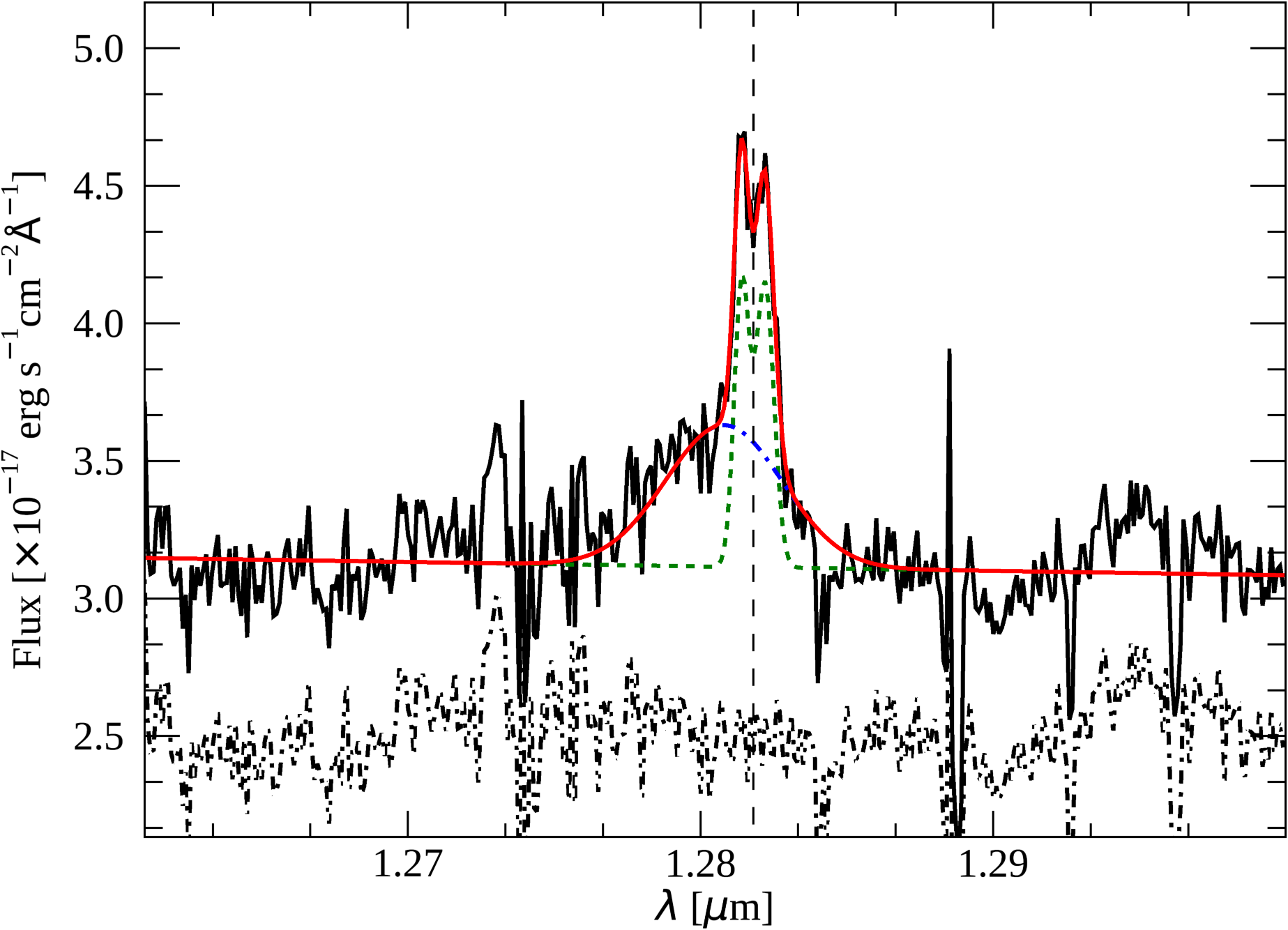}
\endminipage\hfill

\caption{Section of the nuclear spectrum of PKS~B1934-63 showing the {\rm Fe}\,{\rm \scriptsize II}~1.257~$\upmu$m (\textit{top panel}) and the Pa$\beta$ (\textit{bottom panel}) emission lines (black solid line). The best-fit model is shown with the red solid line. The residuals of the fit are normalized and plotted below the spectrum (black dot-dashed line). The vertical dashed lines mark the restframe wavelength of the emission line. For the Pa$\beta,$ the model includes two narrow components (green dotted line) and a broad component (blue dashed line).} 
\label{pabeta_ironlines}
\end{figure}

\end{document}